\documentclass[aps,groupedaddress,showpacs,showkeys,eqsecnum]{revtex4}
\usepackage{amsmath}
\usepackage{graphicx}

\newcommand{\tr}{\text{tr}}

\newcommand{\bfb}{\mbox{\boldmath $b$}}
\newcommand{\bfk}{\mbox{\boldmath $k$}}
\newcommand{\bfp}{\mbox{\boldmath $p$}}
\newcommand{\bfP}{\mbox{\boldmath $P$}}
\newcommand{\bfq}{\mbox{\boldmath $q$}}
\newcommand{\bfv}{\mbox{\boldmath $v$}}
\newcommand{\bfx}{\mbox{\boldmath $x$}}
\newcommand{\bfy}{\mbox{\boldmath $y$}}
\newcommand{\bfz}{\mbox{\boldmath $z$}}
\newcommand{\bfnabla}{\mbox{\boldmath $\nabla$}}
\newcommand{\PP}{{f}}
\newcommand{\KK}{{U}}

\begin{document}

\title{On the cascade approach to the quantum multiscattering problem}

\author{L.L. Salcedo}
\email{salcedo@ugr.es}

\affiliation{
Departamento de F\'{\i}sica Moderna,
Universidad de Granada,
E-18071 Granada, Spain
}

\date{\today} 

\begin{abstract}
The multiscattering problem is studied in the matrix density
formalism. We study how to isolate the quasi-classical degrees of
freedom in order to connect with a cascade approach. The different
problems that arise, as well as their possible solutions, are
discussed and exemplified with a pion-nucleus model.
\end{abstract}


\maketitle

\section{Introduction}
The multiscattering problem is hard to solve in a fully
quantum-mechanical context. Consider, for instance, inclusive
pion-nucleus scattering, which will be our model system. Around the
$\Delta(3,3)$ resonance several channels are open, besides the elastic
one: absorption of the pion, inelastic, single charge exchange and
double charge exchange. Some reactions can take place several times
for the pion inside the nucleus. All the reactions interfere with each
other and their typical reaction probabilities are strongly dependent
on the region of the nucleus. Cascade methods \cite{Snider:1960none}
which reduce the complicated looking output of the reaction to simple
steps, seem appropriate to deal with such a problem. However, cascade
methods often involve drastic semi-classical simplifications not
completely under control. Typically the reaction probabilities are
taken from the free cross section or at most Pauli blocking is
included (and as a consequence are one-body mechanisms) and a
classical propagation is used in between two collisions. Even so, the
potential ability to embody known physics of the problem and the
versatility of the cascade approach makes it interesting to study how
it could be improved to systematically include quantum corrections.

Cascade method have been widely used in the context of heavy ion
collisions, where one wants to describe the dynamical evolution of a
large number of particles. In the crudest approach only two-body
collisions of free particles are considered, without any mean field
effect \cite{Chen:1968none,Cugnon:1982ic}. Complementary to this is
the hydrodynamic approach or the more sophisticated time-dependent
Hartree Fock method, where the mean field is well described but
collisions are neglected \cite{Negele:1981tw,Strottman:1994um}. Both
models are merged into microscopic kinetic or transport models.
Particularly successful has been the Boltzmann-Uehling-Uhlenbeck (BUU)
equation which embodies mean-field effects, two-body collisions and
Pauli blocking \cite{Bertsch:1984gb,Kruse:1985hy,Wolf:1993gg}.
Transport models go back to Boltzmann and are widely used in physics
of plasmas of all kinds \cite{Elze:2002bt}. (See \cite{Elze:1989un,%
Aichelin:1991xy,Blaettel:1993uz,Geiger:1995he,Henning:1995sm,%
Ko:1996yy,Bass:1998ca,Bass:1998vz,Roberts:2000aa} for reviews on the
transport theory approach from different points of view.) In heavy ion
collisions, microscopic transport models are applied not only to
nucleons but also to mesons (in particular pions) and resonances
\cite{Cassing:1990dr}. Transport models have also been used directly
for studying pion-nucleus reactions \cite{Harp:1973ek,%
Ginocchio:1980qb,Fraenkel:1982bm,Cugnon:1988tc,Salcedo:1988md,%
VicenteVacas:1994bk,Engel:1994jh} or other meson-nucleus reactions
\cite{Nara:1997ks}.

In classical kinetic theory the basic quantity is the distribution
function in phase-space, $\PP(\bfx,\bfp,t)$, which satisfies a
transport equation of the type gain-loss as in Eq.~(\ref{eq:1.4})
below. It has long been recognized that the Wigner transform
\cite{Wigner:1932eb,Moyal:1949sk} is the natural way to derive a
quantum transport equation \cite{Yaffe:1982vf,Carruthers:1983fa,%
Aichelin:1991xy,Elze:2002bt}. The subject has been brought to a high
degree of sophistication, as needed for instance in the description of
quark-gluon plasmas, where it has to include thermal effects, be
consistent with relativistic invariance as well as covariance under
non-Abelian gauge transformations \cite{Danielewicz:1984kk,%
Elze:1989un,Henning:1995sm,Zhuang:1996pd,Ochs:1998qj,Blaizot:1999xk}.
The field of quantum transport theory is currently quite active and
with many open issues, both of conceptual and of practical interest
\cite{Elze:2002bt}.


The mark of cascade models is the classical propagation of the
particles between two collisions. This, which at first sight may be
considered as a drawback, could also be one of their main virtues, for
this quasi-classical propagation does indeed exist, furthermore it is
dominant between to well separated successive collisions. Consider,
for instance, the following form of the uncertainty principle
\begin{subequations}
\label{eq:1.1all}
\begin{eqnarray}
\left| \Delta t\right| &\approx & 
\frac{\hbar}{\left| E-E_{\text{on-shell}}
(\bfp)\right|}\,, \label{eq:1.1}\\
\left| \Delta\bfx\right| &\approx & \frac{\hbar}{\left| \bfp - 
\bfp_{\text{on-shell}} (E)\right|} \,. \label{eq:1.2}
\end{eqnarray}
\end{subequations}
These relations show that only on-shell states can propagate a long
time or distance, and that these long lasting states can be taken as
classical in the sense that only they survive as $\hbar$ if formally
taken to zero. Furthermore, from \hbox{Eqs.}  (\ref{eq:1.1all}) one
finds for large $\Delta t,\Delta\bfx$
\begin{equation}
\left| \frac{\Delta\bfx}{\Delta t}\right| \approx
\left| \frac{dE}{d\bfp}\right|_{\text{on-shell}} \,,
\label{eq:1.3}
\end{equation}
which is the classical Hamilton's equation. This crude argument can be
made more precise once a definite prescription is chosen to
simultaneously use conjugate variables such as $(t,E)$ or
$(\bfx,\bfp)$ in the quantum mechanical context, for instance,
Wigner's prescription. Of course, \hbox{Eqs.} (\ref{eq:1.1all}) only
hold as long as interaction does not take place, but they show the
existence of two scales in the multiscattering problem: one due to the
mean free path of classical states and the other due to the
uncertainty principle, {\it i.e.\/}, propagation of virtual
states. Multiscale problems usually make trouble to approaches which
fail to explicitly include such a feature.  It seems more promising to
divide the problem into two parts: first, the quasi-local virtual
states are integrated out in such a way that the elementary vertices
of the microscopic theory are substituted by classical effective
$N$-body quasielastic plus absorption probabilities, and second, these
probabilities are then used in a cascade approach, where only
quasi-classical states show up explicitly.
 
In order to clarify the meaning of the rather abstract program above,
we should place it in the appropriate context. One of our motivations
has been to understand and justify the success of the approach in
\cite{Salcedo:1988md,VicenteVacas:1994bk}. There all the pion-nucleus
inclusive reactions are computed along the following lines: the pion
self-energy is computed in nuclear matter including all the Feynman
graphs considered relevant for energies around the resonance. The
imaginary part of this self-energy is then considered as a pion
``width'' against decay of the elastic channel into reaction
states. With the help of a local density prescription, the total
reaction cross section is then calculated. Furthermore, use is made of
Cutkosky rules \cite{Cutkosky:1960sp,Itzykson:1980bk} in order to
separate the reaction width into the several reaction channels,
namely, absorption and quasielastic (with or without charge
exchange). This information is then used in a Monte Carlo simulation
of the path of the pion inside the nucleus: the pion is treated
classically in between collisions but the reaction probabilities are
computed microscopically from Feynman graphs.  Several questions arise
from this rather intuitive approach, such as how does it follow from a
purely quantum-mechanical calculation?, how can the classical and
local density approximations be systematically improved?  how to avoid
double counting between, say, a genuine three-body absorption
mechanism included in the microscopic calculation and a quasielastic
followed by a two-body absorption coming from the Monte Carlo
simulation, having both the same final state? The Monte Carlo
simulation for the pion can be cast in the form of a transport
equation,
\begin{equation}
\partial_t \PP(\bfx,\bfp,t) = \int d^3x'\,d^3p'\,Q(\bfx
,\bfp;\bfx',\bfp') \PP(\bfx',\bfp',t) -
R(\bfx,\bfp) \PP(\bfx,\bfp,t) \,,
\label{eq:1.4}
\end{equation}
where $Q(\bfx,\bfp;\bfx',\bfp')$ represents the unit time probability
for the transition $(\bfx',\bfp')\to (\bfx, \bfp)$ in phase space and
$R(\bfx,\bfp)$ is the probability of leaving the state
$(\bfx,\bfp)$. $Q$ contains both elastic and quasielastic processes
while $R$ is the total reaction rate. The absorption rate is thus
\begin{equation}
A(\bfx,\bfp) = R(\bfx,\bfp) - \int d^3x'\,d^3p'
\,Q(\bfx',\bfp';\bfx,\bfp)\,.
\label{eq:1.5}
\end{equation}
Then, which are exactly the kernels $R$, $Q$, if any, that will
produce the same cross sections as the Schr\"odinger equation does?
Note that we shall actually deal with just one particle cascading
through the nucleus so $\PP(\bfx,\bfp,t)$ is the probability density
in phase-space rather than the real density as in standard kinetic
theory of plasmas. The equation is however formally identical to a
transport equation.

In this work we address those issues. Our main concern has been
to write exact quantum-mechanical equations in such a way that the
connection with cascade methods were immediate. To this end, a density
matrix formalism is used plus Wigner's prescription, in order to
achieve a well-defined classical limit. The Wigner transformation has
been applied before to study scattering, most notably by Remler in a
series of papers \cite{Remler:1975fm,Remler:1979sj,Remler:1981du,%
Gyulassy:1983pe}. Our emphasis is, however, different
since we are interested in isolating the $N$-body quasielastic and
absorption rates seen by a quasi-classical particle so that the
correct quantum results are recovered in a cascade model.

We shall not be concerned here with relativistic invariance (although
relativistic kinematics is allowed and actually used in applications
involving pions at intermediate energy), thermal effects, or gauge
invariance, however, we shall find that some of the findings in these
more sophisticated fields are also of interest here. We shall find
that a full Wigner transform, in space and also in time, is needed in
order to include inelastic channels, a point not usually realized in
non-relativistic applications of the Wigner transform to collision
theory.

As we have said practical cascade models are often very simplified
from the many-body point of view. Typically, one deals with classical
particles moving in a mean field potential, and classical collisions
using in-vacuum cross sections or decaying width in-vacuum lifetimes,
restricted by Pauli blocking in the final states. This procedure
mimics at a classical level the evolution described by Feynman graphs.
The problem with a direct diagrammatic approach is, of course, that a
realistic description would require to carry out the computation to
graphs of arbitrarily high orders. The cascade method aims at an
efficient procedure to carry out a resummation of those graphs, at the
price of a classical description. To improve on this approach it would
be interesting to make a formulation using exact relations between
resummed sets of graphs, much in the line of the kinetic Kadanoff-Baym
or the Schwinger-Dyson equations, but with an explicit $\hbar$
dependence and within a space-time framework, and then find a
prescription to integrate out the virtual, non-classical, intermediate
states. The improvement of this approach resides in the fact that,
once precise definitions for the effective reaction probabilities are
given, one is no longer constrained to use in-vacuum estimations for
them and, in principle, many-body effects can consistently and
systematically be included. This work should be regarded as an attempt
in this direction. Within a particular simplified model we try to
implement the previous program. In doing so, we introduce ideas, some
of them hopefully new, that presumably will be present in future, more
systematic, developments.

The paper is organized as follows: In Section \ref{sec:II} we show
that indeed knowledge of the evolution of the density matrix in the
Wigner's form, taken as a density of classical projectiles, gives rise
to the correct quantum-mechanical cross section. In Section
\ref{sec:III} the one particle system is studied with emphasis on its
Wigner's form and the classical limit. In Section \ref{sec:IV} the
same analysis is carried out for a particle in an optical potential,
{\it i.e.\/}, for the elastic channel. Section \ref{sec:V} is devoted
to the general many-body evolution equation exemplified with a simple
pion-nucleus model, with only pions and particle-hole $\text{(ph)}$
excitations as physical degrees of freedom. Section \ref{sec:VI} shows
how the ph degrees of freedom can be removed in order to obtain a
purely pionic evolution equation. In Section \ref{sec:VII} the virtual
pionic degrees of freedom are identified and integrated out in order
to obtain effective $N$-body quasielastic rates. In Section
\ref{sec:VIII} the actual outcome of our scheme is illustrated in
simple cases. Finally in Section \ref{sec:IX} we summarize our
conclusions.

\section{Cross section in the simulation approach}
\label{sec:II}

In this Section we shall assume that the proper simulation procedure
has already been carried out (how to do that will be the subject of
subsequent Sections) and our present purpose is to show that in this
case the correct fully quantum mechanical cross section is obtained by
the usual method. We start by relating the $S$-matrix in Wigner's
representation to the cross section, and later contact will be taken
with the time evolution operator which is closer to an actual
simulation procedure.

Let us assume that the Monte Carlo simulation gives us directly the
$\hat\rho_{\text{in}}$, $\hat\rho_{\text{out}}$ relationship
\begin{equation}
\hat\rho_{\text{out}} = \hat S\,\hat\rho_{\text{in}}\,\hat S^\dagger \,,
\label{eq:2.1}
\end{equation}
where $\hat\rho$ is the density matrix and $\hat S$ is the scattering
matrix. Then we shall have
\begin{equation}
\rho_{\text{out}} (u) = \int d^6v \, S(u,v) \rho_{\text{in}} (v)\,.
\label{eq:2.2}
\end{equation}
Here $u$, $v$ are points in the phase space $(\bfx, \bfp)$, $d^6v =
d^3x_vd^3p_v$, $\rho(u)$ is the density matrix in Wigner's form
\cite{Wigner:1932eb,Moyal:1949sk,Yaffe:1982vf,Carruthers:1983fa}
\begin{equation}
\rho(\bfx, \bfp) = \int d^3y\,e^{-i\bfy\cdot \bfp/\hbar}
\textstyle
\left\langle \bfx + \frac{1}{2} \bfy \left| \hat\rho\right| \bfx
- \frac{1}{2} \bfy \right\rangle
\label{eq:2.3}
\end{equation}
and $S(u,v)$ is a real function related to the Wigner's form of $\hat
S$ to be interpreted below as the ``probability'' density of going from
$v$ to $u$ in phase space.
 
The cross section from some initial state $|i\rangle$ to some final state
$|f\rangle$ is (see for instance \cite{Galindo:1991bk})
\begin{equation}
\sigma(i \to f) = \int d^2b \, \Big|\big\langle f\big| \hat S \,
e^{-i{\hat{\bfP}}\cdot {\bfb}/\hbar} \big|i \big\rangle \Big|^2 \,,
\label{eq:2.4}
\end{equation}
$\bfb$ being the impact parameter vector and $\hat{\bfP}$ the
momentum operator. In addition $|i\rangle$, $|f\rangle$ represent {\em
normalized} initial and final states \cite{Galindo:1991bk}. Using the
property of the Wigner's representation
\begin{equation}
\tr\big(\hat X\hat Y\big) = 
\int \frac{d^6u}{\left(2\pi\hbar\right)^3} X(u) Y(u)
\label{eq:2.5}
\end{equation}
the cross section can be rewritten as
\begin{equation}
\sigma(i \to f) = \int d^2 b d^3{x} d^3{p} \frac{d^3{x}' d^3{p}'}
{( 2\pi\hbar )^3}
\rho_f (\bfx', \bfp') \,S(\bfx', \bfp'; \bfx,\bfp)
\rho_i (\bfx - \bfb, \bfp) \,.
\label{eq:2.6}
\end{equation}
Now for $|f\rangle$ we take a plane wave with momentum $\bfp_f$
\begin{equation}
\rho_f (\bfx, \bfp) = \left( 2\pi\hbar\right)^3 
\delta (\bfp - \bfp_f) \,.
\label{eq:2.7}
\end{equation}
This, of course, spoils the dimensional counting, but that will be
fixed later:
\begin{equation}
\sigma(i\to \bfp_f) = \int
d^2b\,d^3x'\,d^3x\,d^3p\,S(\bfx^\prime, \bfp_f;
\bfx+\bfb,\bfp) \rho_i (\bfx, \bfp)\,.
\label{eq:2.8}
\end{equation}
The next step is to use that the fine details of the projectile wave
function are not relevant, {\it i.e.\/} only its momentum distribution
matters. This allows us to rewrite (\ref{eq:2.8}) in the form
\begin{equation}
\sigma(i\to \bfp_f) = \int d^2b\,d^3x^\prime\,d^3p\,
S(\bfx^\prime, \bfp_f;\bfb,\bfp)
\int d^3 x \rho_i (\bfx,\bfp) \,.
\label{eq:2.9}
\end{equation}
Equation (\ref{eq:2.9}) follows from \hbox{Eq.} (\ref{eq:2.8})
provided
\begin{equation}
\bfnabla_{\bfx} \int d^2b\,d^3x^\prime\,
S(\bfx^\prime,\bfp_f; \bfx + \bfb,\bfp) = 0 \,.
\label{eq:2.13}
\end{equation}
Physically this is clearly true: the integration on $\bfb$ projects
out the $\bfx_\perp$ dependence. Furthermore, varying
$\bfx_\parallel$ amounts to a change in the initial position of the
projectile along the same incoming trajectory, but that only shifts
$\bfx^\prime$ which is integrated out. A proof of (\ref{eq:2.13}) is
given in Appendix \ref{app:A}.

Then for $|i\rangle$ normalized but with a narrow momentum
distribution we can take
\begin{equation}
\int d^3x\rho_i (\bfx,\bfp) = \rho_i (\bfp) \approx
\left(2\pi\hbar\right)^3 \delta(\bfp - \bfp_i) 
\label{eq:2.10}
\end{equation}
and so
\begin{equation}
\sigma (\bfp_i \to \bfp_f ) = \left( 2\pi\hbar\right)^3
\int d^2b\,d^3x^\prime\,
S(\bfx^\prime, \bfp_f; \bfb,\bfp_i ) \,.
\label{eq:2.11}
\end{equation}
The correct dimensions are recovered by integrating $\bfp_f$ within
a solid angle $d\Omega_f$
\begin{equation}
\frac{d\sigma( \bfp_i\to {\hat{\bfp}}_f) }{d\Omega_f} = 
\int dp_f\,p^2_f\,d^2b\,d^3x^\prime\,
S( \bfx^\prime, \bfp_f; \bfb,\bfp_i ) \,.
\label{eq:2.12}
\end{equation}
This is the desired relationship between cross section and Monte Carlo
output: to obtain $d\sigma/d\Omega$ several projectiles with random
impact parameter, represented by the integration over $\bfb$, should
be thrown against the target, the simulation procedure (contained in
the function $S ( \bfx^\prime, \bfp_f;\bfb,\bfp_i)$\,) will put them
in $(\bfx^\prime, \bfp_f )$ after the interaction, but only the
scattering angle is relevant (represented by the integration over
$\bfx^\prime$ and $p_f$). $S ( \bfx^\prime,\bfp_f; \bfb,\bfp_i)$ can
then be identified with the density probability of going from $(\bfb,
\bfp_i)$ to $(\bfx^\prime,\bfp_f )$ due to the interaction. It can be
noted, however, that this ``probability density'' is not necessarily
positive. Positivity is only required for the cross section, {\it
i.e.\/}, after integration. Likewise in general, conservation of
energy $\left|\bfp_f\right| = \left|\bfp_i\right|$ is only achieved
after integration over $\bfb$ and $\bfx^\prime$.

Another remark is in order regarding (\ref{eq:2.12}). Unitarity of
$\hat S$ (or also Eq. (\ref{eq:2.2})) implies
\begin{equation}
\int d^6u\,S(u,v) = 1 \,,
\label{eq:2.14}
\end{equation}
then a direct use of (\ref{eq:2.12}) yields
\begin{equation}
\int d\Omega_f \frac{d\sigma( \bfp_i \to {\hat{\bfp}}_f) }{d\Omega_f} 
= \int d^2b = \infty \,.
\label{eq:2.15}
\end{equation}
As usual, this means that the non-interacting part of $\hat S$ must be
removed from the cross section:
\begin{equation}
S(u,v) = Z(u) \,\delta(u-v) + T(u,v) \,,
\label{eq:2.16}
\end{equation}
where $T$ is less singular than $\delta(u-v)$ and represents the
scattering probability while $Z(u)$ is the non-interaction
probability. $T$ instead of $S$ should be used in (\ref{eq:2.12})
(whether $\hat S$ is unitary or not).

The $\hat S$ matrix is a convenient theoretical tool but actually the
simulation procedure is more directly related to the evolution
operator rather than the $\hat S$ matrix: a projectile is sent and the
outgoing distribution is observed much later. This is described by
\begin{equation}
\hat\rho_{t_2} = e^{- i\left( t_2 - t_1\right) \hat H/\hbar}
\,\hat\rho_{t_1} e^{i\left(t_2-t_1\right)\hat H/\hbar}\,; \quad
-t_1,t_2 \to +\infty \,.
\label{eq:2.17}
\end{equation}
In Wigner's form we then have
\begin{equation}
\rho(u,t_2) = \int d^6v\,\KK(u,v;t_2-t_1 )\rho (v, t_1) \,.
\label{eq:2.18}
\end{equation}
It has to be shown that for large $t = t_2-t_1$, $\KK(u,v;t)$ does the
same job as $S(u,v)$ did in (\ref{eq:2.12}). For large enough $t$
\begin{equation}
e^{-i\left(t_2-t_1\right)\hat H/\hbar} \approx e^{-it_2\hat H_0/\hbar}
\hat S e^{it_1\hat H_0/\hbar} = e^{-it\hat H_0 /\hbar} \hat S \,,
\label{eq:2.19}
\end{equation}
where $\hat H_0$, $\hat H$ represent the free and full Hamiltonians
respectively, and we have used the well known property $[\hat H_0,\hat
S]=0$ \cite{Galindo:1991bk,Newton:1982bk}. This allows to relate the
two functions $S(u,v)$ and $\KK(u,v;t)$ as in \hbox{Eq.} (\hbox{A.2}),
namely,
\begin{equation}
\KK(\bfx^\prime,\bfp^\prime;\bfx,\bfp;t) \approx
\int\frac{d^3y d^3q }{(2\pi\hbar)^3}
e^{i\bfq\cdot(\bfx^\prime-\bfy)/\hbar -
i\,\left(H_0 \left(\bfp^\prime +\frac{1}{2}\bfq\right) - H_0
\left(\bfp^\prime - \frac{1}{2}\bfq\right)\right) t/\hbar} 
S (\bfy, \bfp^\prime;\bfx,\bfp) \,.
\label{eq:2.20}
\end{equation}
Now, if $\KK$ is used in (\ref{eq:2.12}) instead of $S$, the $\bfx'$
integration gives $\bfq=0$ and the result is the same, that is,
\begin{equation}
\frac{d\sigma( \bfp_i\to {\hat{\bfp_f}}) }{d\Omega_f} = 
\int dp_f\,p^2_f\,d^2b\,d^3x^\prime\,
\KK(\bfx^\prime, \bfp_f; \bfb,\bfp_i;t) \,.
\label{eq:2.12n}
\end{equation}
Once again, the non-interaction probability is to be removed from
$\KK$. Also note that the result does not depend on $t$ since $\bfq$
vanishes in (\ref{eq:2.20}) after $\bfx'$ integration. This requires
$t$ to be large so that (\ref{eq:2.19}) holds.

Let us summarize the outcome of this Section. A simple-minded method
to obtain the cross section would be: 1) to compute the kernel of the
evolution equation, $\KK(u,v;t)$; 2) to use it in a simulation procedure
as a transition probability density from the point $v$ in phase space
at time $t_1$ to the point $u$ at $t_2$; 3) to count the outgoing
particles to extract the cross section as in a real experiment. What
has been shown here is that the naive expectation is indeed
correct. Note that we use ``simulation procedure'' to mean a method
solving an equation like (\ref{eq:2.18}). The fact that $\KK(u,v;t)$
will not be positive definite in general can be a technical problem
but the equation itself is well-defined.

\section{The single-particle evolution equation}
\label{sec:III}

In this Section we review the simplest case of a one-particle system
and its connection with a classical description.

Let us construct \hbox{Eq.} (\ref{eq:1.4}) for a single-particle
system, with Hamiltonian $\hat H$. We start with the density matrix
evolution equation
\begin{equation}
i\hbar \frac{d\hat\rho}{dt} = \left[ \hat H,\hat\rho\right]\,
\label{eq:3.1}
\end{equation}
and rewrite it using the Wigner transformation, as defined in
(\ref{eq:2.3}), to eventually consider its classical limit. The
product of two operators can be dealt with by means of the identity
\begin{eqnarray}
\big(\hat A\hat B\big) \left( \bfx,\bfp\right) &=& \int
\frac{d^3{y}\,d^3{q} }{ (\pi\hbar)^3} \, \frac{d^3{z}\,d^3{k} }{
(\pi\hbar)^3} A(\bfy, \bfq)\,B(\bfz,\bfk)
e^{i2\left(\bfq - \bfk\right) \cdot \bfx/\hbar}\
e^{i2\left(\bfk-\bfp\right)\cdot \bfy/\hbar}\, 
e^{i2\left( \bfp- \bfq\right)\cdot \bfz/\hbar} \,,
\label{eq:3.2}
\end{eqnarray}
where the left-hand side stands for the Wigner's form of ${\hat
A}{\hat B}$ at $(\bfx,\bfp)$. This can be written more compactly using
the notation (simplectic scalar product)
\begin{eqnarray}
u\wedge v = - v\wedge u &=& \bfx \cdot \bfq - \bfy
\cdot \bfp \,,
\quad
u = \left(\bfx,\bfp\right)\,,
\quad v=\left(\bfy, \bfq\right)\,,
\label{eq:3.4}
\end{eqnarray}
as
\begin{equation}
\big(\hat A\hat B\big)(u) = \int \frac{d^6v }{ (\pi\hbar)^3}\,
\frac{d^6w }{ (\pi\hbar)^3} A(v) B(w)\,e^{i2(v-u)\wedge
(w-u)/\hbar} \,.
\label{eq:3.5}
\end{equation}
In order to study the classical limit, we should transform
this expression into one with better properties as $\hbar$ goes to
zero. To this end we use the identity (in one dimension)
\begin{equation}
e^{ixp/\hbar} = 2\pi\hbar\, e^{i\hbar \partial_x\partial_p} \delta(x)
\delta(p)
\label{eq:3.3}
\end{equation}
where the exponential in the right-hand side is to be expanded as a
series of powers of $\hbar$. This identity can be established by
considering both sides as distributions on $e^{iax}$ as test function,
for arbitrary $a$. It can immediately be extended to any number of
dimensions, and also
\begin{equation}
e^{iu\wedge v/\hbar} = \left(2\pi\hbar\right)^6\,
e^{i\hbar\partial_u\wedge\partial_v} \delta(u) \delta(v) \,.
\label{eq:3.6}
\end{equation}
Then (\ref{eq:3.5}) can be cast in the form
\begin{eqnarray}
\big(\hat A\hat B\big)(u) &=& \int \frac{d^6v }{(\pi\hbar)^3}\,
\frac{d^6w }{ (\pi\hbar)^3} A(u+v) B(u+w) (\pi\hbar)^6\,
e^{\frac{1}{2}i\hbar
 \partial_v\wedge \partial_w} \delta(v) \delta(w) \nonumber\\
&=& e^{\frac{1}{2}i\hbar \partial_v\wedge\partial_w} A(u+v)
B(u+w)\bigg|_{v=w = 0}
\label{eq:3.7}
\end{eqnarray}
or simply \cite{Imre:1967none}
\begin{equation}
\big(\hat A\hat B\big) (u) = e^{\frac{1}{2}i\hbar
\partial_u^{(A)} \wedge \partial_u^{(B)}} A(u) B(u) \,.
\label{eq:3.8n}
\end{equation}
Expanding in powers of $\hbar$ yields
\begin{equation}
\big(\hat A\hat B\big) (u) = 
A(u) B(u) + \frac{i\hbar}{2} \left\{ A(u),B(u)\right\}_P + \cdots \,.
\label{eq:3.8}
\end{equation}
The zeroth order shows that operators commute in the classical limit
and the first correction introduces the usual Poisson bracket, 
$\left\{A,B\right\}_P= \partial A\wedge \partial B$.
 
The evolution equation for $\rho(u,t)$, \hbox{Eq.} (\ref{eq:3.1}),
takes the form:
\begin{subequations}
\label{eq:3.9}
\begin{eqnarray}
i\hbar\frac{\partial\rho(u,t)}{ \partial t} &=& 
\left( e^{\frac{1}{2} i\hbar\partial^{(H)} \wedge 
\partial^{(\rho)}} - e^{\frac{1}{2}i\hbar
\partial^{(\rho)}\wedge \partial^{(H)}} \right) H(u) \rho(u,t)
\label{eq:3.9a} \\
&=& 
\textstyle
2 i \sin \big( \frac{1}{2}\hbar\, \partial^{(H)} \wedge
\partial^{(\rho)}\big)\, H(u)\, \rho(u,t)\,.
\label{eq:3.9b}
\end{eqnarray}
\end{subequations}
At lowest order in $\hbar$ we find
\begin{equation}
\frac{\partial\rho}{\partial t} = \left\{ H, \rho \right\}_P + 
{\cal O}(\hbar^2)\,,
\label{eq:3.10}
\end{equation}
which is the classical equation of evolution in phase space.
This equation is also referred to as the Liouville equation in
mechanics, the Vlasov equation in plasma physics (particularly when
the particles move coupled to an electromagnetic field), or
(collisionless) transport equation in the context of kinetic theory.
Correspondingly, Eq.~(\ref{eq:3.9b}) is the quantum transport
equation. The approach of using the Wigner transformation to derive a
quantum transport equation has become standard
\cite{Carruthers:1983fa} and has been extended in various ways,
including finite temperature \cite{Danielewicz:1984kk,Henning:1995sm}
relativistic treatments \cite{DeGroot:1980dk,Mrowczynski:1994hq},
Abelian and non-Abelian gauge covariant definitions of the Wigner
function as well as second quantization definitions of a Wigner
operator \cite{Elze:1989un}.

For subsequent developments, it will be convenient to consider also
the more general case of a non-Hermitian Hamiltonian $\hat H = \hat
H_R + i\hat H_I$. The new evolution equation is
\begin{equation}
i\hbar \frac{d\hat\rho}{dt} = \hat H\hat\rho - \hat\rho\hat H^\dagger\,,
\label{eq:3.1n}
\end{equation}
and the corresponding equation in Wigner's form would be
\begin{eqnarray}
\partial_t\rho (u,t) &=& \frac{2}{\hbar} \cos\big( {\textstyle \frac{1}{2}}\hbar
\partial^{(H)} \wedge \partial^{(\rho)}\big) H_I (u) \rho(u,t) 
+ \frac{2}{\hbar} \sin \big( {\textstyle \frac{1}{2}}\hbar \partial^{(H)} \wedge
\partial^{(\rho)}\big) H_R (u) \rho(u,t) \nonumber \\
&=& \frac{2}{\hbar} H_I (u) \rho(u,t) +\left\{ H_R(u), \rho(u,t)\right\}_P + 
{\cal O}(\hbar)\,.
\label{eq:3.11}
\end{eqnarray}

Throughout this Section we have used the classical limit in a rather
formal manner. The problem of fixing the $\hbar$ dependence in an
expression is not a trivial or even well-defined one. In practice, our
point of view has been to adopt the Wigner's form of an operator as
that with a smooth classical limit or plainly as that without any
$\hbar$ dependence at all. However, it is clear that this cannot be
true for {\em all} operators \footnote{Assume, for instance, that
$\rho(\bfx,\bfp)$, constructed out of some $\hat\rho$, is non-zero
only in two localized regions of the phase space such that in one of
them it is positive and is negative in the other. The positive region
should be larger than ${\cal O}((2\pi\hbar)^3)$ due to the uncertainty
principle while the negative one can be at most of size ${\cal
O}((2\pi\hbar)^3)$ since $\hat\rho$ is a positive operator. If we were
to reconstruct $\hat\rho$ using a different value for $\hbar$ it could
not be very different in order to satisfy both constraints.}. After
all $\hbar$ is not small, it is rather unity in natural units. The
classical limit should be understood as a physical limit of small
fluctuations. Depending of the system this may correspond to large
times or distances, large momenta or energies, weak coupling (or
sometimes strong coupling), low densities or large number of degrees
of freedom, among others.

\section{Evolution equation and optical potential}
\label{sec:IV}

In this Section, we extend the study of the evolution equation for the
density matrix and its classical limit by including some many-body
effects by means of an optical potential. Let us consider the
following energy-dependent Hamiltonian:
\begin{subequations}
\label{eq:4.1}
\begin{eqnarray}
\hat H(E) &=& \hat H_0 + \hat V_{\text{opt}} (E) \,,
\label{eq:4.1a} \\
\hat V_{\text{opt}} (E)&=&  - \int \frac{dE'}{\pi}\, 
\frac{\hat V_I(E')}{ E-E' + i\eta} \,.
\label{eq:4.1b}
\end{eqnarray}
\end{subequations}
$H_0$ is energy independent and Hermitian, and contains the free and
Hartree-Fock pieces of the particle self-energy. On the other hand,
$\hat V_{\text{opt}} (E)$ is energy dependent and hence non Hermitian
and non instantaneous, and contains the intermediate states not in the
elastic channel. $i\hat V_I$ denotes the anti-Hermitian part of $ \hat
V_{\text{opt}}$. This is the absorptive part of the optical potential
and is non positive definite.

The Schr\"odinger equation in this case takes the form
\begin{equation}
E|E\rangle = \hat H (E) |E\rangle \,,
\label{eq:4.2}
\end{equation}
or in time representation
\begin{equation}
i\hbar\partial_t |\psi,t\rangle = 
\int d\tau\,\hat{h}(\tau)|\psi,t-\tau\rangle\,,
\label{eq:4.3}
\end{equation}
where
\begin{eqnarray}
|\psi,t\rangle &=& 
\int \frac{dE}{2\pi\hbar} e^{-iEt/\hbar} \psi(E) |E\rangle \,, 
\\
\hat{h}(\tau) &=& 
\int \frac{dE}{2\pi\hbar} e^{-iE\tau/\hbar} \hat H(E) 
\nonumber \\
&=& \delta(\tau) \hat H_0+ \frac{i}{\pi\hbar} \theta (\tau) \int
dE\,e^{-iE\tau/\hbar} \hat V_I(E) \,.
\label{eq:4.4}
\end{eqnarray}
The non-locality in time in (\ref{eq:4.3}) comes from non-elastic
intermediate states represented by the optical potential. As in the
previous Section, we would like to construct an evolution equation for
$\hat\rho(t)=|\psi,t\rangle\langle\psi,t|$, however, it is easy to see
that $\hat\rho(t)$ does not satisfy an autonomous equation
\footnote{To see this, consider arbitrarily changing the phases of
$|\psi,t\rangle$ for all times $t$ before some $t_0$. Clearly this
manipulation does not affect $\hat\rho(t)$ for $t<t_0$, however, due
to the non-locality in time in (\ref{eq:4.3}), the evolution of
$|\psi,t\rangle$ and $\hat\rho(t)$ will be modified after $t_0$. This
shows that $\hat\rho(t)$ does not satisfy an autonomous evolution
equation.}.
The problem is that $\hat\rho(t)$ does not actually
contain the same information as $|\psi,t\rangle$ does; it loses track
of the phases.  This would be irrelevant if $|\psi,t\rangle$ where the
wave function of the whole system, but it is not: it only describes
the elastic part. Instead, we have to consider the more general set of
operators
\begin{equation}
\hat\rho(t,t') = |\psi,t\rangle \langle \psi,t'|
\label{eq:4.5}
\end{equation}
as well as their Wigner's transformed (in time-energy)
\begin{equation}
\hat\rho(t,E) = \int d\tau\,e^{iE\tau/\hbar} 
\textstyle
\hat\rho ( t + \frac{1}{2}\tau, t - \frac{1}{2}\tau )
\label{eq:4.6}
\end{equation}
which are Hermitian.
Note that the equal-time density matrix is recovered through energy
integration:
\begin{equation}
\hat\rho(t) = \int \frac{dE}{2\pi\hbar}\hat\rho(t,E) \,.
\label{eq:4.6n}
\end{equation}
Similarly 
we define the Wigner transform of the Hamiltonian 
$\hat H$ (with $\hat H(t,t'):= \hat{h}(t-t')$)
\begin{equation}
\hat H(t,E) = \hat H(E) \,.
\label{eq:4.7}
\end{equation}
In order to write an evolution equation for $\hat\rho(t,E)$, it is
more convenient to start with the energy representation:
\begin{subequations}
\begin{eqnarray}
E|E\rangle &=& \hat H(E) |E\rangle  \,,
\\
\hat\rho (E_1, E_2) &=& |E_1\rangle \langle E_2|
\label{eq:4.8}
\end{eqnarray}
\end{subequations}
so that
\begin{subequations}
\label{eq:4.9all}
\begin{eqnarray}
E_1\hat\rho (E_1, E_2) &= \hat H(E_1) \hat\rho(E_1, E_2) \,,
\\
E_2 \hat\rho(E_1, E_2) &= \hat\rho (E_1, E_2) \hat H^\dagger (E_2) \,.
\label{eq:4.9}
\end{eqnarray}
\end{subequations}
Subtracting both equations and using the notation $E = \frac{1}{2}
(E_1 +E_2)$, $\omega = E_1-E_2$ and $\hat\rho(\omega,E) =
\hat\rho(E_1, E_2)$, we obtain
\begin{equation}
\omega\hat\rho(\omega,E) = 
\textstyle
\hat H ( E + \frac{1}{2} \omega )
\hat\rho (\omega, E) - \hat\rho(\omega,E) \hat H^\dagger  ( E - \frac{1}{2}
\omega )\,,
\label{eq:4.10}
\end{equation}
which can also be written in terms of $\hat\rho(t,E)$ by Fourier transforming
$\omega$:
\begin{eqnarray}
i\hbar\partial_t \hat\rho(t,E) &=& e^{\frac{1}{2}i\hbar
\partial^{(\rho)}_t \partial^{(H)}_E } \hat H (E) \hat\rho (t,E) 
- e^{-\frac{1}{2}i\hbar \partial^{(\rho)}_t \partial^{(H)}_E} \hat\rho (t,E)
\hat H^\dagger(E) \,.
\label{eq:4.11}
\end{eqnarray}
This is the evolution 
or {\em transport equation}
in presence of an optical potential. It
is an autonomous equation for $\hat\rho(t,E)$ as a function of $t$
because, as a consequence of energy conservation, no energy derivatives
of $\hat\rho$ appear in it. Note that if we take the ``classical''
limit in the right-hand side by taking the explicit $\hbar\to 0$, Eq.
(\ref{eq:3.1n}) is recovered. 

Eq. (\ref{eq:4.11}) has been written using a partial Wigner form,
namely, in time-energy. It can be further expanded by using the full
space-time Wigner's transformation (in time-energy and
position-momentum). This yields the following more symmetrical form
of the transport equation which generalizes Eq.~(\ref{eq:3.9a})
\begin{eqnarray}
i\hbar\partial_t\rho(u) &=& e^{-\frac{1}{2}i\hbar \partial^{(H)}_u
\wedge \partial^{(\rho)}_u} H(u) \rho(u) 
- e^{\frac{1}{2}i\hbar\partial^{(H)}_u \wedge \partial^{(\rho)}_u}
\rho(u) H^\dagger (u) \,,
\label{eq:4.12}
\end{eqnarray}
where now $u$ denotes $(t,\bfx; E,\bfp)$, and $u\wedge v = tE' -
Et' - \bfx\cdot \bfp' + \bfp\cdot \bfx'$.
This formula holds actually for a general non conservative and non
instantaneous Hamiltonian $\hat H(t,E)$. In what follows we will not,
in general, expand the formulas by expliciting the position-momentum
part of the Wigner transform.

As it stands, (\ref{eq:4.11}) is of little usefulness since it is
non-local in time and so all the time derivatives of $\hat\rho$ will
contribute to the first one. Fortunately, the expansion in powers of
$\hbar$ can be used to bypass this undesirable feature. First let us
expand \hbox{Eq.} (\ref{eq:4.11}) using the separation $\hat H(E) = \hat
H_0+\hat V_R(E) + i\hat V_I(E)$,
\begin{eqnarray}
i\hbar\partial_t \hat\rho &=& \left[ \hat H_0+\hat V_R,\hat\rho\right]_- +
\left[ i \hat V_I, \hat\rho\right]_+ \nonumber \\
&&+ \left[ \partial_E \hat V_R, \frac{i\hbar}{2}\partial_t \hat\rho\right]_+ +
\left[ i \partial_E \hat V_I, \frac{i\hbar}{2} \partial_t \hat\rho\right]_- \nonumber \\
&&+ \frac{1}{2} \left[ \partial^2_E \hat V_R, \Big( \frac{i\hbar}{2}\Big)^2
\partial^2_t \hat\rho\right]_-
+ \frac{1}{2} \left[ i \partial^2_E \hat V_I, \Big( \frac{i\hbar}{2}\Big)^2
\partial^2_t \hat\rho\right]_+ + \cdots \,,
\label{eq:4.13}
\end{eqnarray}
where $\left[\,,\,\right]_-, \left[\,,\,\right]_+$ stand for
commutator and anticommutator respectively. Then, let us demand that
there exist a classical limit at all for $\partial_t\rho(\bfx, \bfp;
t,E)$. The first term in the right-hand side of (\ref{eq:4.13}) poses
no problem because it is of order $\hbar$ due to the commutator
(cf. (\ref{eq:3.8})). Analogously the third and higher order terms
have explicit $\hbar$ in them. So we will require
\begin{equation}
\hat V_I = {\cal O}(\hbar)
\label{eq:4.14}
\end{equation}
for $\partial_t\rho$ to exist in the limit $\hbar\to 0$. Note that
this implies $\hat V_R = {\cal O}(\hbar)$ too, due to (\ref{eq:4.1b}).
Under this assumption, it follows that the first and the second terms
in \hbox{Eq.} (\ref{eq:4.13}) are ${\cal O}(\hbar)$. The third term is
${\cal O}(\hbar^2)$. The fourth term is ${\cal O}(\hbar^3)$, and the
others are ${\cal O}(\hbar^4)$ and ${\cal O}(\hbar^3)$. In general,
higher order time derivatives of $\hat\rho$ appear only at higher
order in $\hbar$. This allows to express $\partial_t\hat\rho$ in terms
of $\hat\rho(t,E)$ only (without time derivatives) at any given order
in $\hbar$: $\partial_t\hat\rho$ in the third and fourth terms are
eliminated using (\ref{eq:4.13}) recursively, $\partial^2_t\hat\rho$
in the fifth and sixth terms are eliminated by applying
$(i\hbar\partial_t)$ once to (\ref{eq:4.13}), etc. Explicitly, through
second order, we obtain
\begin{eqnarray}
i\hbar\partial_t\hat\rho &=& \left[\hat H_0,\hat\rho\right]_- +
i\left[ \hat V_I,\hat\rho\right]_+ \nonumber \\
&&+ \left[ \hat V_R,\hat\rho\right]_- + \frac{1}{2}\left[
\partial_E\hat V_R,\left[\hat H_0,\hat\rho\right]_- + i\left[ \hat V_I,
\hat\rho\right]_+ \right]_+ \nonumber \\
&&+ \, {\cal O}(\hbar^3) \,.
\label{eq:4.15}
\end{eqnarray}
This expansion is expected to be only asymptotic. In what follows we
will not, in general, explicitly expand the equations to put them in a
manifestly instantaneous form, as done here, but this procedure can be
carried out if needed.

It is worth noticing that the real part of the optical potential $\hat
V_R$, only appears at higher order than the free Hamiltonian $\hat
H_0$ or the absorptive part of $\hat V_I$. In many cascade
calculations only $\hat V_I$ is used, ({\em i.e.}, the cross
section). (\ref{eq:4.15}) indicates that this is a kind of classical
approximation.
 
As we have seen, the relation (\ref{eq:4.14}) is needed in order for
\hbox{Eq.} (\ref{eq:4.15}) to make sense. Its origin is clearer if
$\hat V_{\text{opt}}$ is considered as a self-energy in many-body
language \cite{Fetter:1974bk}. From \hbox{Eq.} (\ref{eq:4.1}) the
Hartree pieces are included in $\hat H_0$ and then $\hat V_{\text{opt}}$
contains only self-energy pieces with loops, the only diagrams with
imaginary part. As shown in \cite{Coleman:1973jx}, each loop gives a
further power in $\hbar$ to a diagram, consistently with \hbox{Eq.}
(\ref{eq:4.14}).

Let us consider the propagation of a particle in infinite nuclear
matter. Due to translational invariance, all relevant operators are
functions of the momentum and commute among them (internal degrees of
freedom are neglected here). In this case (\ref{eq:4.15}) becomes
\begin{equation}
\partial_t \rho(t,E) = - \frac{\Gamma(E)}{\hbar} \rho(t,E)
\label{eq:4.16}
\end{equation}
where $\Gamma(E) = {\cal O}(\hbar)$ is the width and is given as a
power series in $\hbar$. A closed form is more easily obtained
directly from (\ref{eq:4.10})
\begin{equation}
\textstyle
- i\Gamma(E) = H ( E-\frac{1}{2}i\Gamma ) 
- H^\dagger ( E +\frac{1}{2}i\Gamma ) \,.
\label{eq:4.17}
\end{equation}
From here, expanding in $\hbar$, the well-known quasiparticle result
\cite{Fetter:1974bk}
\begin{subequations}
\begin{eqnarray}
\Gamma(E) &\approx & - 2 Z(E) V_I(E) \,, 
\\
Z(E) &=& \left( 1 - \partial_E V_R (E)\right)^{-1} 
\label{eq:4.18}
\end{eqnarray}
\end{subequations}
appears in order $\hbar^2$.

It can be noted that from the two equations (\ref{eq:4.9all}) we have
extracted only one equation, namely, (\ref{eq:4.10}). Taking the
semi-sum instead of the difference in (\ref{eq:4.9all}) one obtains
\begin{eqnarray}
E \hat\rho(t,E) &=& \frac{1}{2}e^{\frac{1}{2}i\hbar
\partial^{(\rho)}_t \partial^{(H)}_E } \hat H (E) \hat\rho (t,E)
+\frac{1}{2} e^{-\frac{1}{2}i\hbar \partial^{(\rho)}_t \partial^{(H)}_E} \hat\rho (t,E)
\hat H^\dagger(E) \,,
\label{eq:4.11n}
\end{eqnarray}
which is a kind of energy-shell {\em constraint equation}
\cite{Elze:1989un,Elze:2002bt}. At lowest order in $\hbar$, in the
full Wigner form, it just says that $E=H(u)$, i.e., the density matrix
is on-shell. By construction this equation is consistent with the
transport equation (\ref{eq:4.11}). In fact, using again an asymptotic
expansion in $\hbar$, we can put this equation in instantaneous
form. The lowest orders are
\begin{eqnarray}
E\hat\rho &=& \frac{1}{2}\left[\hat H_0,\hat\rho\right]_+ 
+ \frac{1}{2}\left[ \hat V_R,\hat\rho\right]_+ \nonumber \\
&& +\frac{i}{2}
\left[ \hat V_I,\hat\rho\right]_- 
+ \frac{i}{4}\left[
\partial_E\hat V_I,\left[\hat H_0,\hat\rho\right]_- +i\left[ \hat V_I,
\hat\rho\right]_+ \right]_+ \nonumber \\
&&+ \, {\cal O}(\hbar^3) \,.
\label{eq:4.15n}
\end{eqnarray}
If the transport and constraint equations are written as
$\partial_t\hat\rho=L_t\hat\rho$ and $E\hat\rho=L_E\hat\rho$, with the
linear actions $L_t$ and $L_E$ defined by (\ref{eq:4.15}) and
(\ref{eq:4.15n}), respectively, the compatibility amounts to the
statement $[L_t,L_E]=0$, which can be verified also by explicit
calculation to the order shown.

As often emphasized, the two equations, transport and constraint, are
needed for a proper description of the evolution \cite{Elze:1989un,
Henning:1995sm,Zhuang:1996pd,Ochs:1998qj,Elze:2002bt}. In next
sections we will also make use of two equations, however, our approach
will involve a different set of kinematic equations, (\ref{eq:5.13})
and (\ref{eq:5.15}), or (\ref{eq:6.11}) and (\ref{eq:6.13})). This is
further discussed at the end of Section \ref{sec:VII}.

An equal-time density matrix is often used in non-relativistic
transport equations \cite{Remler:1975fm}. In the relativistic case the
equal-time formulation can also be used
\cite{Bialynicki-Birula:1991tx} but Lorentz invariance is only
manifest by using the space-time Wigner transform \cite{Elze:2002bt}.
As discussed in \cite{Zhuang:1996pd,Ochs:1998qj} both formulations are
equivalent without introducing further approximations, the space-time
formulation being nevertheless richer since it contains the energy
distribution \cite{Zhuang:1996pd,Ochs:1998qj}.

The equivalence of the two approaches (equal-time vs. time-energy
Wigner distribution functions) holds whenever the Hamiltonian is
instantaneous, that is, whenever $H(u)$ is independent of $E$. This
follows from integrating over the energy in (\ref{eq:4.12}) (a
procedure named energy averaging in \cite{Zhuang:1996pd,Ochs:1998qj}),
recalling the relation (\ref{eq:4.6n}) between the two density
matrices, and noting that $\partial_E$ acts only on $\rho(u)$ (it is
not necessary to assume that the Hamiltonian is conservative, $H(u)$
may depend on $t$). The situation for a non instantaneous dynamics is
different and it requires the use of the space-time form, as done
here. This is not related to relativity but rather to the fact that
the particle can leave the elastic channel and spend some time in
other inelastic intermediate states before returning to the elastic
Hilbert space. In this case to derive an autonomous evolution equation
for the equal-time density matrix is no longer straightforward. We
will deal with a closely related problem in Section \ref{sec:VII}.


\section{Non-elastic states}
\label{sec:V}

\begin{figure}
\includegraphics[scale= 0.5]{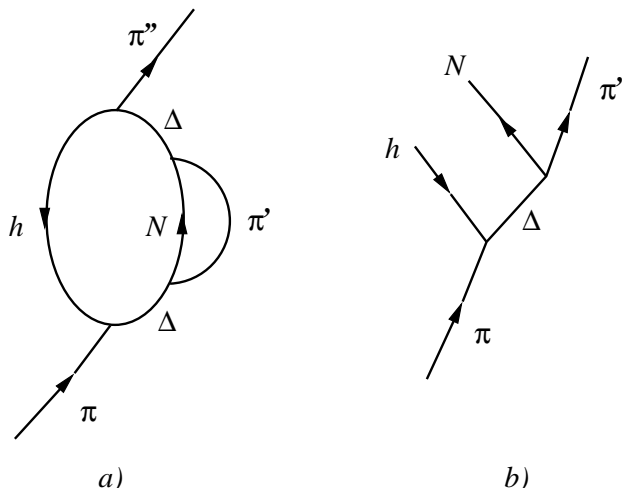}
\caption{Typical graphs in $\Delta$-hole model: $a$) Contribution to
the pion self-energy; $b$) a quasielastic process.}
\label{fig:1}
\end{figure}

In the previous Section we studied the time evolution of the density
matrix describing the elastic part of the wave function. Here we
would like to study how to describe the non-elastic part. To fix
ideas consider the scattering of pions by nuclei in the
$\Delta$-isobar region \cite{Brown:1975di}. Typical processes are
those depicted in \hbox{Fig.} \ref{fig:1}.
\hbox{Fig.} \ref{fig:1}$a$ shows a self-energy graph contributing to
the pion optical potential: the incoming pion $\pi$ collides with a
nucleon of the nucleus which is excited to a $\Delta$-isobar and
leaves a hole $h$ in the Fermi sea of nucleons, the $\Delta$ couples
further with a $N\pi'$ state. If these intermediate states are only
virtual, {\it i.e.\/}, for a short time of order $\hbar$, the pion
$\pi''$ emerges with the same energy as $\pi$: it is an elastic
scattering and it only contributes to the real part of the optical
potential. On the other hand, if the intermediate particles $\pi'Nh$
are near their mass-shell, the $|\pi'Nh\rangle$ state can live a long
time and we have instead the process in \hbox{Fig.} \ref{fig:1}$b$: it
is a real decay of a pionic mode into pion-particle-hole and
contributes to the imaginary part of the optical, {\it i.e.\/}, gives
a width to the ``elastic'' incoming pion $\pi$.
 
\begin{figure}
\includegraphics[scale= 0.5]{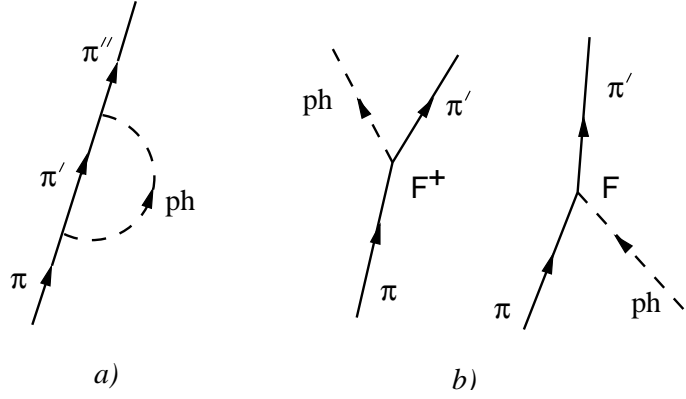}
\caption{The same processes as in \hbox{Fig.} \ref{fig:1}, for the model in
\hbox{Eq.} (\ref{eq:5.1}).}
\label{fig:2}
\end{figure}

We wish to describe such quasielastic reactions (\hbox{Fig.}
\ref{fig:1}$b$) by means of a density matrix formalism appropriate to
connect with cascade calculations. In order not to unnecessarily
obscure the discussion we shall use a simplified model with two kinds
of particles: ``pions'' and ``ph'' (particle-hole), both bosons,
without explicit $\Delta$-isobar or isospin degrees of freedom. In
addition, the equations here will not include pion absorption. The
absorption mechanism is developed in Appendix \ref{app:D} and added
later. The model is given by the following Hamiltonian
\begin{equation}
\hat H = \hat H^{(\pi)}_0 + \hat H^{\text{(ph)}}_0 + \hat H_I 
:= \hat H_f + \hat H_I
\label{eq:5.1}
\end{equation}
where $\hat H^{(\pi)}_0$, $\hat H^{\text{(ph)}}_0$ are one-body operators for pions
and  ph, and $\hat H_I$ is the interaction vertex $\pi\pi\text{ph}$:
\begin{eqnarray}
\hat H_I &=& \int
d^3{x}\,d^3{y}\,d^3{z}\, F(\bfx;\bfy,\bfz) \hat\phi_{\text{ph}} (\bfx)
\hat\phi^\dagger_\pi(\bfy)\hat\phi_\pi(\bfz) + \hbox{h.c.} \nonumber \\
&:=& \hat F + \hat F^\dagger \,,
\label{eq:5.2}
\end{eqnarray}
and as usual
\begin{eqnarray}
\left[ \hat\phi_\pi(\bfx), \hat\phi^\dagger_\pi(\bfy)\right] &=& 
\left[ \hat\phi_{\text{ph}} (\bfx), \hat\phi^\dagger_{\text{ph}} (\bfy) \right] 
= \delta(\bfx-\bfy)\,, \nonumber \\
\left[ \hat\phi_\pi , \hat\phi_{\text{ph}} \right] &=&
\left[ \hat\phi_\pi,\hat\phi^\dagger_{\text{ph}}\right] = 0
\,.
\label{eq:5.3}
\end{eqnarray}
The interaction vertex is assumed to be elementary ({\it i.e.\/},
instantaneous and without energy dependence) but not necessarily
local. Within this schematic model, the diagrams of
\hbox{Fig.} \ref{fig:1} are now described by those in \hbox{Fig.}
\ref{fig:2}.

At $t=-\infty$ the state consists of a single incoming pion. In this
model the number of pions is conserved by all pieces of the
Hamiltonian. On the other hand, $\hat H_f$ preserves the number of
ph's but $\hat F^\dagger$ and $\hat F$ act creating and deleting one
ph, respectively. A generic state of the system will thus contain
exactly one pion plus a number $k$ of ph particles, with $0\le k \le
A$ where $A$ is the mass number of the nucleus
\begin{equation}
|\psi\rangle = \sum^A_{k=0} |\pi,\text{(ph)}^k\rangle 
:= \sum^A_{k=0} |k\rangle \,.
\label{eq:5.5}
\end{equation}
$|\pi\rangle=|k=0\rangle$ corresponds to the elastic channel, and
$|k\rangle$ is obtained after $k$ inelastic steps. Note that in this
model the ph does not have self-energy graphs (unless absorption is
included by a vertex $\pi\text{ph}$ as in Appendix \ref{app:D}). The
conservation of the number of pions and the free propagation of the
ph's are both consequences of the fact that the model does not
implement crossing symmetry for the pions (there are no anti-pions in
this model). In this sense it has some resemblance with the Lee model
\cite{Lee:1954iq,Baz:1969bk} where the number of possible graphs is
severely limited due to the existence of very restrictive conservation
laws on the number of particles. In our case the number of graphs is,
however, considerably larger since the number of ph's is not
restricted: in the one-pion subspace the more general graph consists
of the continuous pion line with zero or more outgoing ph lines
stemming from it plus zero or more ph internal lines with the ph
emitted and reabsorbed by the pion in any order. Since, within the
model, a ph cannot couple to a pion-anti-pion pair there are no ph
self-energy graphs. This model is devised to describe, in a simplified
manner, the problem of pion-nucleus reactions at energies around
resonance or below where pion production is below threshold or barely
so. When we introduce absorption in Appendix \ref{app:D} the pion will
couple to a single ph (modeling the absorption of a virtual pion by a
nucleon). In this case pions and ph can be transmuted into each other
and the number of pions needs not be conserved. (Nevertheless in this
region of energies, states with two or more pions can only be virtual
and we shall simplify further the exposition by not including them in
the formulas.)

Our starting point is the Schr\"odinger equation
\begin{equation}
i\hbar \frac{d}{dt} |\psi,t\rangle = \hat H|\psi,t\rangle \,.
\label{eq:5.4}
\end{equation}
Using that the states $|k,t\rangle$ are linearly independent, we can write
\begin{equation}
i\hbar\frac{d}{dt} |k,t\rangle = \hat H_f |k,t\rangle + \hat F^\dagger|k-1,t\rangle +
\hat F |k+1,t\rangle\,,\quad {k=0,\ldots,A} \,.
\label{eq:5.6}
\end{equation}
(Here and in what follows we use the convention that quantities with
indices $k$ out of the physical range $0\le k \le A$ vanish
identically. In the present case the terms with $|k=-1\rangle$ and
$|k=A+1\rangle$ in the equations for $k=0$ and $k=A$, respectively,
are absent.) Although correct, these equations show an unwanted
symmetry under time reversal: as the pion goes scattering through the
nucleus, the number of ph produced will increase, then we would prefer
that states with higher number of ph's had lower ones as a source and
not conversely. Let us see how to achieve this in the simple case of
$A=2$. Using an energy representation for (\ref{eq:5.6}):
\begin{eqnarray}
E|0,E\rangle &=& \hat H_f |0,E\rangle + \hat F|1,E\rangle \nonumber \\
E|1,E\rangle &=& \hat H_f |1,E\rangle + \hat F^\dagger|0,E\rangle + \hat F|2 ,E\rangle
\label{eq:5.7} \\
E|2,E\rangle &=& \hat H_f |2,E\rangle + \hat F^\dagger|1,E\rangle \,.
\nonumber
\end{eqnarray}
From the last equation $|2,E\rangle = \left( E - \hat H_f +
i\eta\right)^{-1}\hat F^\dagger |1,E\rangle$. The time reversal symmetry is
broken by choosing $+i\eta$. Substituting in the second equation and
using the same method in the resulting equation for $|1,E\rangle$, we
end up with
\begin{eqnarray}
\textstyle   
E|0,E\rangle &=& \hat H_f |0,E\rangle + \hat F \Big( E -
\hat H_f - \hat F \left( E - \hat H_f + i\eta \right)^{-1} \hat F^\dagger +
i\eta \Big)^{-1}\hat F^\dagger|0,E\rangle \nonumber \\
E|1,E\rangle &=& \hat H_f | 1,E\rangle + \hat F \left( E - \hat H_f +
i\eta\right)^{-1} \hat F^\dagger|1,E\rangle + \hat F^\dagger |0,E\rangle \label{eq:5.8} \\
E|2,E\rangle &=& \hat H_f |2,E\rangle + \hat F^\dagger |1,E\rangle \,.
\nonumber
\end{eqnarray}
The new equations have the desired form; the equation of motion of
$|k\rangle$ involves only the states $|k\rangle$ and $|k-1\rangle$. In
general the equations are
\begin{equation}
E|k,E\rangle = \hat H_k (E) |k,E\rangle + \hat F^\dagger|k-1,E\rangle
\label{eq:5.9}
\end{equation}
with
\begin{subequations}
\begin{eqnarray}
\hat H_k(E) &=& \hat H_f + \hat F \hat G_{k+1} (E) \hat F^\dagger \,,
\label{eq:5.10} \\
\hat G_k(E) &=& \left( E - \hat H_k(E) + i\eta\right)^{-1} \,. 
\label{eq:5.11}
\end{eqnarray}
\end{subequations}
Comparing (\ref{eq:5.9}) with (\ref{eq:4.2}), we can see that the
operator $\hat H_k (E)$ plays the role of an optical Hamiltonian for
the state $|\pi\text{(ph)}^k\rangle$. The operators $\hat G_k(E)$ are
the corresponding propagators and in this model they do not contain
intermediates states with less than $k$ ph particles. From our
previous convention $\hat H_{k=A}(E) = \hat H_f$ and $\hat G_{k=A}(E)$
is just the free propagator. On the other hand, $\hat G_{k=0}(E)$ is
the full pion propagator and $\hat H_{k=0}(E)$ is the full pion
optical potential. For this elastic channel the last term in
(\ref{eq:5.9}) is absent and this equation coincides with
(\ref{eq:4.2}).

Upon Fourier transform in time-energy Eq. (\ref{eq:5.9}) provides the
differential-like evolution equation. It can also be written in
integral form as
\begin{equation}
|k,E\rangle = \hat G_k(E)\hat F^\dagger|k-1,E\rangle 
\qquad (k>0) \,.
\label{eq:5.9n}
\end{equation}
Both forms of the evolution equation will be used subsequently.

To proceed, we define the density matrices
\begin{equation}
\hat\rho_k(E_1,E_2) := |k,E_1\rangle\langle k,E_2|
\label{eq:5.12}
\end{equation}
which satisfy (using (\ref{eq:5.9n}))
\begin{equation}
\hat\rho_k(E_1,E_2) = \hat G_k(E_1) \hat F^\dagger\, \hat\rho_{k-1} (E_1,E_2) 
\hat F \,\hat G^\dagger_k(E_2) \qquad (k>0) \,.
\label{eq:5.13}
\end{equation}
In Wigner's representation, this equation gives $\hat\rho_k(t,E)$ in
terms of $\hat\rho_{k-1}(t,E)$. In order to obtain the
differential-like equation for $\hat\rho_k(t,E)$ we first rewrite
\hbox{Eq.} (\ref{eq:5.13}) as
\begin{equation}
E_1\hat\rho_k(E_1,E_2) = \hat H_k(E_1) \hat\rho_k(E_1,E_2) + \hat F^\dagger \hat\rho_{k-1}
(E_1,E_2) \hat F \hat G^\dagger_k(E_2)
\label{eq:5.14}
\end{equation}
and with the same notation and method used to obtain (\ref{eq:4.10})
and (\ref{eq:4.11}), we have
\begin{eqnarray}
\omega\hat\rho_k(\omega,E) &=& 
\textstyle
\hat H_k ( E + \frac{1}{2}\omega ) 
\hat\rho_k (\omega,E) - \hat\rho_k(\omega,E) 
\hat H^\dagger_k (E - \frac{1}{2}\omega ) 
\nonumber \\
&& 
\textstyle
- \hat G_k (E + \frac{1}{2}\omega ) 
\hat F^\dagger \hat\rho_{k-1} \hat F 
+ \hat F^\dagger \hat\rho_{k-1} \hat F 
\hat G^\dagger_k ( E-\frac{1}{2}\omega ) \,,
\label{eq:5.15}
\end{eqnarray}
or in time representation
\begin{eqnarray}
i\hbar\partial_t\hat\rho_k(t,E) &=& e^{\frac{1}{2}i\hbar
\partial^{(\rho)}_t \partial^{(H)}_E} \hat H_k (E) \hat\rho_k (t,E) -
e^{-\frac{1}{2}i\hbar \partial^{(\rho)}_t \partial^{(H)}_E} \hat\rho_k (t,E)
\hat H^\dagger_k(E) \nonumber \\
&&- e^{\frac{1}{2}i\hbar \partial^{(\rho)}_t \partial^{(G)}_E} \hat G_k (E)
\hat F^\dagger \hat\rho_{k-1} (t,E) \hat F 
+ e^{-\frac{1}{2}i\hbar \partial^{(\rho)}_t \partial^{(G)}_E} 
\hat F^\dagger \hat\rho_{k-1} (t,E) \hat F \hat G^\dagger_k(E) \,.
\label{eq:5.16}
\end{eqnarray}
This can be regarded as an extension of \hbox{Eq.} (\ref{eq:4.11}) to
account for the creation of the new $|k\rangle$ states out of
$|k-1\rangle$ when $k>0$. Using (\ref{eq:5.10}) it is readily verified that this equation
preserves unitarity, namely,
\begin{equation}
\partial_t \sum^A_{k=0} \tr \left(
\hat\rho_k (t,E) \right) = 0 \,.
\label{eq:5.17}
\end{equation}
 
To summarize this Section: we have been able to write an evolution
equation for the quantities $\hat\rho_k (t,E)$, namely Eq.
(\ref{eq:5.16}), which in full Wigner's form provides us with the
relationship between simulation-like quantities, $\hat\rho_k
(\bfx,\bfp;t,E)$, on one hand, and microscopic-like quantities, the
Green's functions $\hat G_k(E)$ and optical Hamiltonians $\hat
H_k(E)$, on the other, without semiclassical approximations involved.
However, this equation is not fully satisfactory because too much
information is contained in $\hat\rho_k$, namely, all nuclear degrees
of freedom as well as far from classical (highly virtual) pionic
degrees of freedom. We deal with such problems in next Sections.

\section{Removal of ph degrees of freedom}
\label{sec:VI}

The quantities $\hat\rho_k(t,E)$ and their evolution equations contain
information both on ``pions'' and on `` ph '' states. If we are only
interested in the pionic reactions it is convenient to simplify the
problem by just working with the pionic degrees of freedom. This is
the approach in \cite{Salcedo:1988md}, where the simulation only
traces the path of the pion inside of the nucleus. This requires to
eliminate the ph degrees of freedom.
 
\begin{figure}
\includegraphics[scale= 0.5]{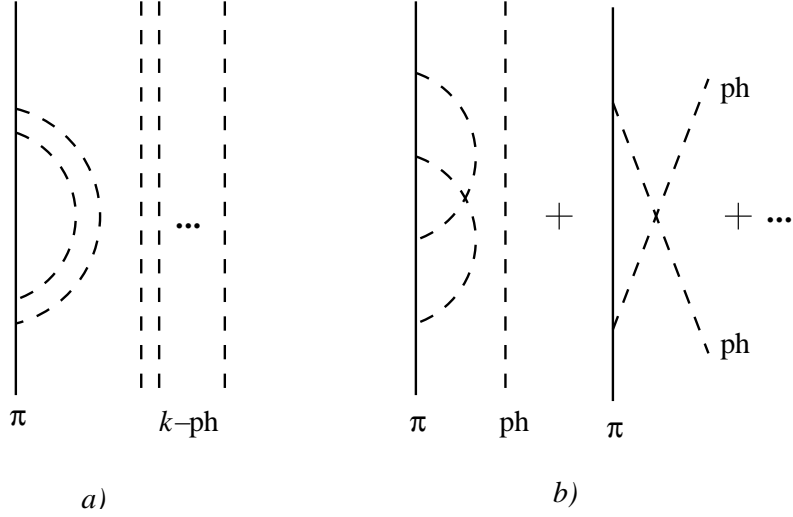}
\caption{$a$) Typical graph contributing to the directed part of $\hat G_k$; $b$) crossed
  graphs in $\hat G_k$ (for $k=1$).}
\label{fig:3}
\end{figure}

The first idea is to define a new density matrix out of $\hat\rho_k$
for the pion only by taking trace over the ph part. If we attempt to
do so in \hbox{Eq.} (\ref{eq:5.13}), or in the other equations, we
find that this new operator $\tr_{\text{ph}} \left(\hat\rho_k\right)$
does not obey an autonomous set of equations; the knowledge of
$\tr_{\text{ph}}\left(\hat\rho_{k-1}\right)$ does not provide us with
$\tr_{\text{ph}}\left(\hat\rho_k\right)$ because this information is
only partial. On the other hand, $\tr_{\text{ph}}$ does not remove the
ph energies which are included in $E$, and in addition, the time $t$
is a common time for the pion and the ph's and that may not be the
most appropriate choice. It seems thus necessary to disentangle the
different energy and time dependences in $\hat\rho_k$ in order to find
a density matrix truly depending only on pionic energy and time, as
well as $\bfx_\pi$ and $\bfp_\pi$. Likely, the problem is not trivial,
and in fact I have only partially succeeded in solving it: there is a
solution if the propagators $\hat G_k(E)$ only contain direct graphs,
as those in \hbox{Fig.}  \ref{fig:3}$a$, and no crossed terms,
\hbox{Fig.} \ref{fig:3}$b$. In this case the ph particles are
distinguishable and $1,2,\ldots,k$ labels the order in which the ph's
have been produced.

When the particles are distinguishable, the total wave function at
time $t$ is just the product of individual wave functions taken at the
common time $t$. In energy this corresponds to a convolution.
Mathematically we have
\begin{equation}
i\hbar\hat G_k(E) = \big[i\hbar \hat G^{(k)}_\pi \circ i\hbar\hat G^{(1)}_{\text{ph}} \circ
i\hbar\hat G^{(2)}_{\text{ph}} \circ \cdots \circ i\hbar\hat G^{(k)}_{\text{ph}}\big](E) + \text{(crossed
terms)}
\label{eq:6.1}
\end{equation}
where the explicit term is that associated to direct graphs, and we
have introduced the following notation: the symbol ``$\circ$'' stands
for convolution over the energy dependence
\begin{equation}
\big[ A\circ B\big](E) = \int \frac{dE_A}{2\pi\hbar}\, \frac{dE_B}{2\pi\hbar} 2\pi\hbar
\,\delta(E-E_A-E_B) A(E_A) B(E_B)  \,,
\label{eq:6.2}
\end{equation}
the operator $\hat G^{(\ell)}_{\text{ph}}(E)$ is the free propagator of the
$\ell$-th ph (as noted previously, the ph particles do not have
self-energy graphs in this model)
\begin{equation}
\hat G^{(\ell)}_{\text{ph}} (E) = \left( E- \hat H^{(\ell)}_{\text{ph}} +
i\eta\right)^{-1}\,,
\label{eq:6.3}
\end{equation}
($\hat H^{(\ell)}_{\text{ph}}$ is the free Hamiltonian of the $\ell$-th
ph, included in $\hat H_0^{\text{(ph)}}$). Finally, the operator $\hat
G^{(k)}_\pi(E)$ is the pion propagator in presence of $k$ ph's, but
including direct self-energy graphs only. Thus it is given
recursively by (cf. Fig. \ref{fig:3}$a$)
\begin{subequations}
\label{eq:6.4}
\begin{eqnarray}
\hat G^{(k)}_\pi(E) &=& \left( E - \hat H^{(k)}_\pi (E) + i\eta\right)^{-1} \,,
\label{eq:6.4a} \\
\hat H^{(k)}_\pi (E) &=& \hat H^{(\pi)}_0 - i\hbar \hat F 
\big[\hat G^{(k+1)}_\pi \circ \hat G^{(k+1)}_{\text{ph}}\big](E) \hat F^\dagger
\label{eq:6.4b}
\end{eqnarray}
\end{subequations}
where $\hat H^{(k)}_\pi (E)$ is the optical Hamiltonian of the pion in
presence of $k$ ph's and the operators $\hat F,\hat F^\dagger$ act only
on the $k+1$-th ph. Note that $\hat G^{(\ell)}_{\text{ph}}$ and $\hat
G_\pi$ act in the $\ell$-th ph and pionic Hilbert spaces respectively
and they commute.

The purpose of selecting the direct graphs was to be able to
disentangle the energy-time and position-momentum degrees of freedom
carried by each individual particle in the state $\hat\rho_k(t,E)$.
This is achieved as follows. As shown in detail in Appendix
\ref{app:B}, if crossed terms are dropped in $\hat G_k$, a new (more
detailed) state $|E_\pi, E_1,\ldots,E_k\rangle_k$ can be defined,
depending on $\pi$ and ph degrees of freedom $(\bfx$ and $\bfp)$ and
also on their energies, which is related to the state $|k,E\rangle$ in
\hbox{Eq.}  (\ref{eq:5.9}) by means of
\begin{equation}
|k,E\rangle = \int \frac{dE_\pi}{2\pi\hbar}\, \frac{dE_1}{2\pi\hbar} \cdots
\frac{dE_k}{2\pi\hbar} 2\pi\hbar \,\delta\left( E - E_\pi - E_1 - \cdots
-E_k\right) |E_\pi, E_1,\ldots ,E_k\rangle_k \,.
\label{eq:6.5}
\end{equation}
The new state satisfies the integral-like equation, which is analogous
to (\ref{eq:5.9n}),
\begin{equation}
|E_\pi,E_1,\ldots, E_k\rangle_k
= i \hbar \hat G^{(k)}_\pi (E_\pi) i\hbar \hat G^{(k)}_{\text{ph}} (E_k) 
\frac{1}{i\hbar}
\hat F^\dagger 
| E_\pi + E_k, E_1, E_2,\ldots, E_{k-1} \rangle_{k-1} \qquad (k>0) \,.
\label{eq:6.6}
\end{equation}
Here $\hat F^\dagger$ acts on the pion in $|\ \rangle_{k-1}$ and creates the
$k$-th ph, thus building the subset of direct graphs only (and so
$|E_\pi,E_1,\ldots,E_k\rangle_k$ is not symmetric under exchange of
ph's). Actually this equation is used recursively in Appendix
\ref{app:B} to define the states $|E_\pi,E_1,\ldots, E_k\rangle_k$
starting from the $k=0$ state.

Next we define the associated density matrix
\begin{eqnarray}
&& \hat\rho_k( \omega_\pi,E_\pi, \omega_1,E_1,\ldots,\omega_k,
E_k ) := \nonumber \\
&&
\textstyle
\quad
\left|E_\pi + \frac{1}{2}\omega_\pi, E_1 + \frac{1}{2}\omega_1,\ldots, E_k + \frac{1}{2}\omega_k \right\rangle_k
\left\langle E_\pi - \frac{1}{2}\omega_\pi, E_1-\frac{1}{2}\omega_1,\ldots,
E_k - \frac{1}{2}\omega_k \right|_k
\label{eq:6.7}
\end{eqnarray}
from which $\hat\rho_k(\omega,E)$ can easily be recovered by making
use of \hbox{Eq.} (\ref{eq:6.5}). This new density matrix satisfies
the following recurrence equation (which is a translation of
(\ref{eq:6.6}))
\begin{eqnarray}
&&\hat\rho_k (\omega_\pi, E_\pi,\omega_1,E_1,\ldots,\omega_k, E_k ) = 
\textstyle
\hbar^2 \hat G^{(k)}_\pi ( E_\pi + \frac{1}{2}\omega_\pi ) 
\, \hat G^{(k)}_{\text{ph}} ( E_k + \frac{1}{2}\omega_k ) \,\hat F^\dagger
\nonumber \\
&&
\textstyle
\quad \times \hat\rho_{k-1} ( \omega_\pi + \omega_k, E_\pi + E_k, \omega_1,
E_1,\ldots, \omega_{k-1}, E_{k-1} ) \,
\hat F 
\, \hat G^{(k)\dagger}_{\text{ph}} ( E_k - \frac{1}{2}\omega_k ) 
\, \hat G^{(k)\dagger}_\pi ( E_\pi - \frac{1}{2}\omega_\pi ) \,.
\label{eq:6.8}
\end{eqnarray}
This equation represents a definite improvement over \hbox{Eq.}
(\ref{eq:5.13}) because now the trace can be taken over ph degrees of
freedom and a closed set of equations is obtained: on the right-hand
side the trace factorizes in the form $\left\{ 1,\ldots, k-1\right\}
\left\{ k\right\}$ and as a consequence $\tr_{\text{ph}}\hat\rho_k$ is
given in terms of $\tr_{\text{ph}}\hat\rho_{k-1}$.

In order to obtain a pionic density matrix, let us call it
$\hat\rho^{(k)}(\omega_\pi, E_\pi;E)$, the ph energies $E_\ell$ can
easily be integrated out. Also, the ``times'' $\omega_\ell$ should be
fixed. After Fourier transforming, each $\omega_\ell$ becomes the time
$t_\ell$ at which the $\ell$-th ph is detected. In order to preserve
unitarity at any ``pionic'' time $t$, we should have $t$
smaller that any ph time, so we are lead to choose all the ph times
$t_\ell$ as $+\infty$ in the definition of the purely pionic density
matrix, that is,
\begin{eqnarray}
\hat\rho^{(k)} (t,E_\pi;E)
&:=& \lim_{\{t_\ell\}\to + \infty}
\int \frac{dE_1}{2\pi\hbar}\, \frac{dE_2}{2\pi\hbar} \cdots \frac{dE_k}{2\pi\hbar}
2\pi\hbar\,\delta ( E - E_\pi - E_1- \cdots- E_k ) \nonumber \\
&&\quad \times \tr_{\text{ph}} 
\hat\rho_k (t, E_\pi; t_1, E_1,\ldots, t_k, E_k ) \,.
\label{eq:6.9}
\end{eqnarray}
(Where $t$ refers to the pionic time associated to the pionic
frequency $\omega_\pi$. Also we
remark that we are taking the limit of large ph times, and not
integrating over those times.) This is the density matrix in pionic
space describing the pions which have scattered $k$ times (producing
$k$ ph's). In particular,
\begin{equation}
\hat\rho^{(k=0)}(t, E_\pi; E ) =
2\pi\hbar\,\delta(E_\pi-E)\, \hat\rho_{k=0} (t, E)
\label{eq:6.10}
\end{equation}
is the density matrix for the elastic channel.
Clearly, in $\hat\rho^{(k)}(t,E_\pi;E)$, $E$ corresponds to the total
energy, which is conserved. On the other hand $E_\pi$ is the energy
carried by the pion after $k$ collisions and both quantities coincide
in the elastic channel. We further discuss on the coexistence of these
two energies in $\hat\rho^{(k)}(t,E_\pi;E)$ over the end of Section
\ref{sec:VII}.

As shown in detail in Appendix \ref{app:C}, if the definition in
(\ref{eq:6.9}) is used in \hbox{Eq.} (\ref{eq:6.8}), the following
recurrence is found for $\hat\rho^{(k)}$ (analogous to
(\ref{eq:5.13})):
\begin{eqnarray}
\hat\rho^{(k)} (\omega_\pi,E_\pi;E) &=& 
\textstyle
\hat G^{(k)}_\pi (E_\pi + \frac{1}{2}\omega_\pi ) \nonumber \\
&&\quad \times \int \frac{dE'_\pi}{2\pi\hbar} \tr_{\text{ph}} 
\left\{ 
2\pi\hbar\, \delta ( E'_\pi - E_\pi - \hat H_{\text{ph}} ) 
\hat F^\dagger \hat\rho^{(k-1)} (\omega_\pi,E'_\pi;E) \hat F
\right\} \nonumber \\
&&
\textstyle
\quad \times  \hat G^{(k)\dagger}_\pi (E_\pi - \frac{1}{2}\omega_\pi)
,\qquad(k>0)
\label{eq:6.11}\,.
\end{eqnarray}
Moreover, there it is also shown that unitarity is preserved, as a
direct mathematical consequence of taking all $t_{\text{ph}}$ as $+\infty$,
that is,
\begin{equation}
\partial_t \left[ 
\sum^A_{k=0} \tr \int \frac{dE_\pi}{2\pi\hbar} 
\hat\rho^{(k)}( t, E_\pi; E )
\right]  = 0 \,.
\label{eq:6.12}
\end{equation}

To obtain the Schr\"odinger-like equation associated to the
integral-like equation in (\ref{eq:6.11}), we follow the same
procedure as that used to obtain \hbox{Eq.} (\ref{eq:5.15}). This
gives
\begin{eqnarray}
\omega_\pi\hat\rho^{(k)} (\omega_\pi, E_\pi; E )
&=&
\textstyle
\hat H^{(k)}_\pi ( E_\pi + \frac{1}{2}\omega_\pi) 
\, \hat\rho^{(k)} ( \omega_\pi, E_\pi; E )
- \hat\rho^{(k)} (\omega_\pi, E_\pi; E )
\, \hat H^{(k)\dagger}_\pi ( E_\pi - \frac{1}{2}\omega_\pi) \nonumber \\
&& \quad
- \tr_{\text{ph}} \int \frac{dE'_\pi}{2\pi\hbar}\,2\pi\hbar\,\delta (E'_\pi
- E_\pi - \hat H_{\text{ph}}) \nonumber \\
&&
\textstyle
\quad \times \Big[ \hat G^{(k)}_\pi ( E_\pi + \frac{1}{2}\omega_\pi )
\, \hat F^\dagger \, \hat\rho^{(k-1)} ( \omega_\pi, E'_\pi; E )\, \hat F 
\nonumber \\ &&  \textstyle \qquad
- \hat F^\dagger \,\hat\rho^{(k-1)} ( \omega_\pi,E'_\pi; E ) \,
\hat F \, \hat G ^{(k)\dagger}_\pi ( E_\pi - \frac{1}{2}\omega_\pi )
\Big] \,
 \label{eq:6.13}
\end{eqnarray}
and then, after Fourier transforming in $\omega_\pi$,
\begin{equation}
\partial_t\hat\rho^{(k)} (t, E_\pi; E )
= \Big(\partial_t\hat\rho^{(k)} \Big)^{(+)} (t, E_\pi; E )
+ \Big(\partial_t\hat\rho^{(k)} \Big)^{(-)} (t, E_\pi; E )
\label{eq:6.13n1}
\end{equation}
with
\begin{eqnarray}
\Big(i\hbar\, \partial_t\hat\rho^{(k)} \Big)^{(+)} (t, E_\pi; E )
&=&
\textstyle
e^{\frac{1}{2}i\hbar\partial_t^{(\rho)}\partial_{E_\pi}^{(H)}}
\hat H^{(k)}_\pi ( E_\pi ) 
\, \hat\rho^{(k)}(t,E_\pi;E)
\nonumber \\ &&  \quad
- e^{-\frac{1}{2}i\hbar\partial_t^{(\rho)}\partial_{E_\pi}^{(H)}}
\hat\rho^{(k)}(t,E_\pi;E) \,
\hat H^{(k)\dagger}_\pi ( E_\pi ) 
\label{eq:6.13n2}
\end{eqnarray}
and
\begin{eqnarray}
\Big(i\hbar\, \partial_t\hat\rho^{(k)} \Big)^{(-)} (t, E_\pi; E )
&=&
- \tr_{\text{ph}} \int \frac{dE'_\pi}{2\pi\hbar}\,2\pi\hbar\,
\delta (E'_\pi - E_\pi - \hat H_{\text{ph}}) 
\nonumber \\ && \quad \times 
\Big[ 
e^{\frac{1}{2}i\hbar\partial_t^{(\rho)}\partial_{E_\pi}^{(G)}}
\hat G^{(k)}_\pi ( E_\pi)
\, \hat F^\dagger \, \hat\rho^{(k-1)}(t,E'_\pi;E) \, \hat F 
\nonumber \\ &&  \textstyle \qquad
- e^{-\frac{1}{2}i\hbar\partial_t^{(\rho)}\partial_{E_\pi}^{(G)}}
\hat F^\dagger\,\hat\rho^{(k-1)}(t,E'_\pi;E) \,
\hat F \, \hat G ^{(k)\dagger}_\pi(E_\pi)
\Big] \,.
 \label{eq:6.13n3}
\end{eqnarray}
Equations (\ref{eq:6.11}) and (\ref{eq:6.13}) (or equivalently,
(\ref{eq:6.13n1}))
are the relevant result of this Section. In $\partial_t\hat\rho^{(k)}$
we have distinguished two contributions. The first one,
$(\partial_t\hat\rho^{(k)})^{(+)}$, is related to $\hat\rho^{(k)}$
itself and describes a pion (of class $k$) propagating with optical
Hamiltonian $\hat H_\pi^{(k)}$, (cf. Eq. (\ref{eq:4.11})): it contains
both, the ``elastic'' propagation of the pion inside the nucleus, and
the quasielastic steps in which the pion of class $k$ becomes of class
$k+1$. In this sense it can be called the {\it annihilation} part of
$\partial_t\hat\rho^{(k)}$. The second contribution,
$(\partial_t\hat\rho^{(k)})^{(-)}$, accounts for the quasielastic
steps of the form $k-1 \to k$. From the point of view of
$\hat\rho^{(k)}$ it is the {\it creation} part of
$\partial_t\hat\rho^{(k)}$. The Dirac delta in this term indicates
that the ph's are on-shell. Mathematically, this is a direct
consequence of having chosen all the ph times as $+\infty$ in the
definition of the pionic density, (\ref{eq:6.9}).
 
\begin{figure}
\includegraphics[scale= 0.5]{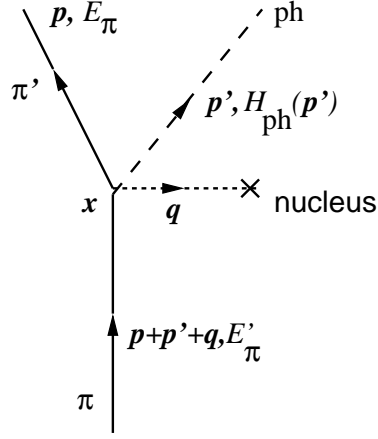}
\caption{Quasielastic process in a finite system.}
\label{fig:4}
\end{figure}

Before studying Eq. (\ref{eq:6.13n1}) in more detail, let us see how it
works in a simple case:
\begin{equation}
\hat F = \int d^3{x}\,g(\bfx)\,\hat\phi_{\text{ph}}(\bfx)
\hat\phi^\dagger_\pi (\bfx)\,\hat\phi_\pi(\bfx)
\label{eq:6.14}
\end{equation}
and $\hat H^{(\pi)}_0$, $\hat H_{\text{ph}}$ functions of the momentum
only. Let us first put the creation part of
$\partial_t\hat\rho^{(k)}$ in Wigner's form.  In a first step
\begin{eqnarray}
&&\Biggl[ \tr_{\text{ph}}\int \frac{dE'_\pi}{2\pi\hbar} \,
2\pi\hbar\,\delta( E'_\pi- E_\pi - \hat H_{\text{ph}}) \,
\hat F^\dagger \,\hat\rho^{(k-1)} ( t,E'_\pi;E) \,\hat F \Biggr]
( \bfx_\pi, \bfp_\pi ) \nonumber \\
&& \qquad = \int \frac{d^3{p}'}{(2\pi\hbar)^3} \, \frac{d^3{q}}{(2\pi\hbar)^3} 
\, g^2 (\bfx_\pi,\bfq) \,
\rho^{(k-1)}(\bfx_\pi,\bfp_\pi+\bfp'+\bfq;t,E_\pi+H_{\text{ph}}(\bfp');E)
\label{eq:6.15}
\end{eqnarray}
where
\begin{equation}
g^2 (\bfx,\bfq) :=
\int d^3{y}\,e^{-i\bfy \cdot \bfq/\hbar}
\textstyle
 g^*(\bfx + \frac{1}{2}\bfy) g(\bfx-\frac{1}{2}\bfy )
\label{eq:6.16}
\end{equation}
and $\bfq$ is the momentum transferred to the nucleus (\hbox{Fig.} \ref{fig:4}).

Next, we shall retain only the leading order in $\hbar$. In this way
several simplifications take place in $(\partial_t\rho)^{(-)}$: i) the
exponentials
$\exp({\pm\frac{1}{2}i\hbar\partial_t^{(\rho)}\partial_{E_\pi}^{(G)}})$
are unity at leading order in $\hbar$; ii) the pion propagator becomes
free, $\hat G_\pi(E)\to ( E - \hat H^{(\pi)}_0 + i\eta )^{-1}$,
because each loop in its self-energy gives a $\hbar$ factor
(\hbox{Eq.}  (\ref{eq:4.14})); iii) in the classical limit the
operators commute (cf. (Eq. (\ref{eq:3.8})); and, iv) using the
identity (\ref{eq:3.3})
\begin{equation}
g^2 (\bfx,\bfq)
\underset{{\hbar\to 0}}{\longrightarrow}
\left(2\pi\hbar\right)^3\,\delta(\bfq) g^2 (\bfx) \,.
\label{eq:6.17}
\end{equation}
After some algebra, the creation part can be written as
\begin{eqnarray}
&& \left(\partial_t \rho^{(k)}\right)^{(-)} ( \bfx, \bfp; t,E_\pi;E) 
= \int \frac{d^3{x}'\,d^3{p}'}{ \left(2\pi\hbar\right)^3}\
\frac{dE'_\pi}{ 2\pi\hbar}\, \tilde Q ( \bfx, \bfp, E_\pi; \bfx', \bfp', E'_\pi) 
\,\rho^{(k-1)} ( \bfx', \bfp'; t,E'_\pi;E) + {\cal O}(\hbar) 
\label{eq:6.18}
\end{eqnarray}
with
\begin{eqnarray}
&& \tilde Q \left( \bfx, \bfp, E_\pi; \bfx', \bfp', E'_\pi\right)
= \frac{1}{ \hbar^2} g^2 (\bfx) \,\delta ( \bfx - \bfx' ) 
2\pi\hbar\,\delta \big( E_\pi - H^{(\pi)}_0 (\bfp)\big)
2\pi\hbar\,\delta (E'_\pi - E_\pi - H_{\text{ph}} ( \bfp' - \bfp )) \,.
\label{eq:6.19}
\end{eqnarray}
We can see that the quasielastic probability in this approximation is
positive, local ($\bfx=\bfx'$) and momentum is conserved $(\bfq = 0$).
The ph are on-shell ($E'_\pi - E_\pi = H_{\text{ph}} ( \bfp' - \bfp
)$) and moreover the outgoing pions are also created on-shell. As
noted in the Introduction the latter fact is a typical consequence of
taking $\hbar\to 0$, due to the uncertainty principle. Mathematically,
the factor $2\pi\hbar\,\delta \big( E_\pi - H^{(\pi)}_0 (\bfp)\big)$
follows from the imaginary part of the free pion propagator $\hat
G_\pi(E)-\hat G_\pi^\dagger(E)$ in (\ref{eq:6.13n3}). Summarizing, the
probability of quasielastic $\tilde Q$ obtained at leading order in
$\hbar$ coincides with the result that would follow from carrying out
a standard nuclear matter calculation ($g =g_c$ constant) in lowest
order in perturbation theory, plus a local density prescription
($g_c\to g(x)$ at the end of the nuclear matter calculation)
\cite{Salcedo:1988md}.
This latter fact is quite remarkable since usually the local density
approximation is put in by hand, and here it follows naturally as
a semiclassical approximation.

The annihilation part of $\partial_t\hat\rho^{(k)}$ comes from
computing the pion optical potential up to one loop and keeping the
free and the imaginary parts (lowest order in $\hbar$ in \hbox{Eq.}
(\ref{eq:4.15})). This gives
\begin{eqnarray}
\left(\partial_t \rho^{(k)}\right)^{(+)} ( \bfx, \bfp; t, E_\pi;E)
&=& \left\{ H^{(\pi)}_0 (\bfp),\rho^{(k)} (\bfx, \bfp; t, E_\pi; E) \right\}_P 
- \tilde R ( \bfx, \bfp; E_\pi) \,\rho^{(k)} ( \bfx, \bfp; t, E_\pi; E)
+ {\cal O}(\hbar)
\label{eq:6.20}
\end{eqnarray}
where
\begin{equation}
\tilde R (\bfx,\bfp;E_\pi) = \frac{1}{ \hbar^2} g^2 (\bfx) 
\int \frac{d^3{p}'}{ \left( 2\pi\hbar\right)^3} 
\, 2\pi\hbar\,\delta \big( E_\pi - H_{\text{ph}} ( \bfp - \bfp') - H^{(\pi)}_0 (\bfp')\big) \,.
\label{eq:6.20n}
\end{equation}
The first term of the right-hand side of Eq. (\ref{eq:6.20}) describes
a classical free propagation of the pion. The second term indicates a
reaction probability rate given by $\tilde R(\bfx,\bfp,E_\pi)$.
Unitarity is verified since
\begin{equation}
\tilde R (\bfx,\bfp;E_\pi) = \int \frac{d^3{x}'\,d^3{p}'}{ \left( 2\pi\hbar\right)^3} 
\, \frac{dE'_\pi}{ 2\pi\hbar} \, \tilde Q (\bfx', \bfp', E'_\pi; \bfx,\bfp, E_\pi) \,.
\label{eq:6.21}
\end{equation}

Of course, the right-hand side of (\ref{eq:6.18}) should be set to 0
for the elastic channel, $k=0$, and similarly, the reaction term is
not present in (\ref{eq:6.20}) when no ph remains in the nucleus,
$k=A$. Otherwise, $\tilde Q$ and $\tilde R$ are independent of $k$ and
$E$ in this approximation. Also note that because the pions are
on-shell the $E_\pi$ dependence can be dropped and \hbox{Eqs.}
(\ref{eq:6.18}) and (\ref{eq:6.20}) have the form of \hbox{Eq.}
(\ref{eq:1.4}), except that the classical propagation has been
extracted from the quasielastic rate $Q$.

Unfortunately, beyond leading order in $\hbar$, Eq. (\ref{eq:6.13n1})
displays two undesirable features: first, the dependence of
$\hat\rho^{(k)}$ on $E_\pi$, which implies that virtual as well as
real pions coexist in $\hat\rho^{(k)}$. We deal with this problem in
next Section. And second, the non locality in time: in general
$\partial_t\hat\rho^{(k)}$ will depend on the previous history of
$\hat\rho^{(k)}$ and $\hat\rho^{(k-1)}$. The instantaneous equations
can be obtained by a method similar to that used for
Eq. (\ref{eq:4.15}): $\partial_t\hat\rho^{(k)}$ is given as a function
of $\partial_t^n\hat\rho^{(k)}$ and $\partial_t^{n'}\hat\rho^{(k-1)},\
n,n' = 0,1, \ldots$, each time derivative carrying a factor
$\hbar$. Then, higher order derivatives on the right-hand side can be
eliminated in terms of lower ones.  For instance, writing
Eq. (\ref{eq:6.13n1}) with an obvious schematic notation, we have for
$k=1,2$
\begin{subequations}
\label{eq:6.22}
\begin{eqnarray}
\partial_t\rho^{(2)} &=&
 \sum^\infty_{n=0} \hbar^n\,N_n^{(+)}\,\partial_t^n\rho^{(2)} +
 \sum^\infty_{n=0} \hbar^n\,N_n^{(-)}\,\partial_t^n\rho^{(1)}  \\
\partial_t\rho^{(1)} &=&
 \sum^\infty_{n=0} \hbar^n\,{N'}_n^{(+)}\,\partial_t^n\rho^{(1)} +
 \sum^\infty_{n=0} \hbar^n\,{N'}_n^{(-)}\,\partial_t^n\rho^{(0)} \,.
\label{eq:6.22a}
\end{eqnarray}
\end{subequations}
Then, to first order in $\hbar$,
\begin{eqnarray}
\partial_t\rho^{(2)} = &&
 \left\{ N_0^{(+)} + \hbar N_1^{(+)}N_0^{(+)} + {\cal O}(\hbar^2) \right\}
\rho^{(2)}  \nonumber \\
&& + \left\{ N_0^{(-)} + \hbar N_1^{(-)}{N'}_0^{(+)} + {\cal O}
(\hbar^2) \right\}\rho^{(1)} \nonumber \\
&& + \left\{ \hbar N_1^{(-)}{N'}_0^{(-)} + {\cal O}(\hbar^2) \right\}\rho^{(0)} \,.
\label{eq:6.23}
\end{eqnarray}
We observe that the equations can be written in an instantaneous form
at the price of expressing $\partial_t\rho^{(k)}$, not only in terms
of $\rho^{(k)}$ and $\rho^{(k-1)}$, but using all densities
$\rho^{(k')}$ with $k'\leq k$. We shall see that a similar phenomenon
occurs when the $E_\pi$ dependence is removed form $\hat\rho^{(k)}$.

\section{Integration of the virtual pions}
\label{sec:VII}

In the previous Section we have succeeded in writing an equation, Eq.~(\ref{eq:6.13n1}), for
the temporal evolution of the density matrix of pions which have
scattered $k$ times, with only pionic degrees of freedom. However, it
is not a fully satisfactory one due to its dependence in the pion
energy, $E_\pi$. Consider, for instance, the evolution of a pion after
its creation ({\it i.e.\/} after the scattering): two frequencies are
relevant, first that characteristic of the source, represented by
$E_\pi$, and second that of the free evolution,
$H^{(\pi)}_0(\bfp_\pi)$. Due to interference, after traveling some
wavelengths only on-shell pions, those with $E_\pi = H^{(\pi)}_0
(\bfp_\pi)$, will survive, unless another scattering takes place. Such
a collapse of the wave function can also be regarded as a classical
limit: if $\hbar$ were very small any time interval would be large
compared with a few periods of the order of $\hbar/ |E_\pi -
H^{(\pi)}_0 (\bfp_\pi)|$ provided that the particle is off-shell. Then
we expect the following relation to hold:
\begin{equation}
\hat\rho^{(k)}(t,E_\pi;E)
\underset{\hbar\to 0}{\longrightarrow} 2\pi\hbar\,
\delta(E_\pi - \hat H_0^{(\pi)})
\label{eq:7.1}
\end{equation}
up to a factor. This relation 
follows from the energy-shell constraint Eq.~(\ref{eq:4.11n}) and is
checked in next Section (cf. Eq.  (\ref{eq:8.12})).

\begin{figure}
\includegraphics[scale= 0.5]{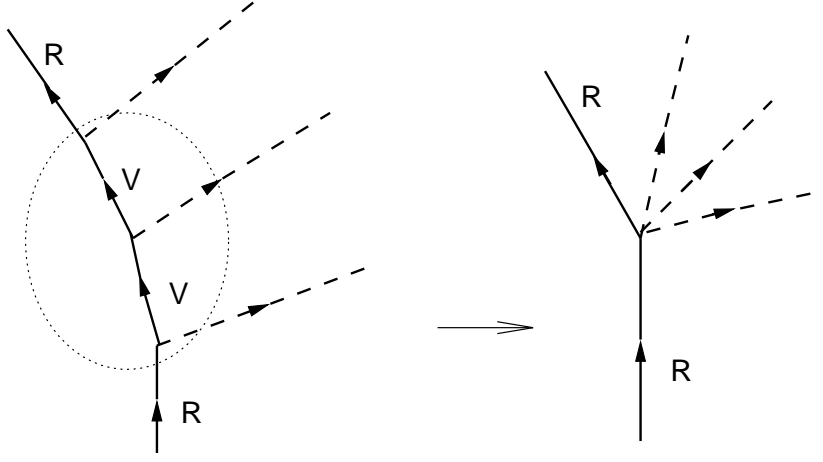}
\caption{Shrinking of virtual intermediate states ($V$) to produce an
effective $N$-body quasielastic step between real pions ($R$).}
\label{fig:5}
\end{figure}

Initially (in the incoming pion beam) the pion is on-shell. After
several quasielastic steps the pion leaves the nucleus and is again
on-shell since only on-shell pions can travel long distances to the
detector. Between two successive collisions the pion can be (nearly)
on-shell (i.e., a {\em real} pion) or off-shell (a {\em virtual}
pion). Consider a typical path of the pion through the nucleus, for
instance $RVVRVR$ (time increasing from right to left) where $R$
stands for real and $V$ for virtual pion, in this case with five
quasielastic collisions. Each collision is of the type one-body
mechanism, producing one ph. The virtual pions will travel a short
distance and the Monte Carlo simulation only needs to trace the real
ones. Within a classical limit, the real pions would follow a
classical trajectory, whereas the path followed by the virtual pions
shrinks. The situation is thus better described as a path $RRR$, that
is, involving real pions only and two collisions. The first collision
ejects two ph, a two-body quasielastic mechanism, whereas the second
collision involves a three-body quasielastic mechanism (see
\hbox{Fig.}  \ref{fig:5}).

Advancing results of the next Section, we can see the different
properties of real and virtual pions as follows. Consider the
classical limit directly in \hbox{Eq.} (\ref{eq:6.11}). By neglecting
the $\pm\omega_\pi/2$ (which carries an $\hbar$ upon Fourier transform
to time representation) in $\hat G_\pi$ and going to time
representation it turns out that $\hat\rho^{(k)}(t,E_\pi;E)$ is
related to $\hat\rho^{(k-1)}(t,E'_\pi;E)$, {\it i.e.\/}, the number of
pions of class $k$ is proportional to the number of pions of class
$k-1$ {\em at the same time}. This is only possible if they live
during a very short period of time, and this is correct for virtual
pions. Of course, for on-shell pions this would be an absurd
consequence which is avoided because for them taking
$\pm\omega_\pi/2\to 0$ is not justified even in the classical limit:
the quadratic $G_\pi G^\dagger_\pi$ divergence (at the on-shell point) is too strong. (This is also the reason for needing differential-like
evolution equations, as (\ref{eq:6.13}) which have softer divergences
than the integral-like (\ref{eq:6.11}).) The virtual pions are not
only irrelevant to final cross sections but their short living times
can only be achieved by interference which is very hard to reproduce
by a simulation. Also, their existence implies that
$\hat\rho(t,E_\pi;E)$ should be strongly non-positive definite.

Our goal is thus to integrate out the virtual pions and end up with an
equation including only real pions. Because the real pions are
on-shell their energy dependence is fixed and its corresponding
density, to be denoted ${\hat \PP_k} (t,E)$, does not depend on
$E_\pi$. In addition, in order to preserve unitarity at each time, and
not only in the $t\to+\infty$ limit, we are led to the following
definition for the density of on-shell pions,
\begin{equation}
{\hat \PP_k} (t,E) := \int \frac{dE_\pi}{2\pi\hbar} \hat\rho^{(k)}
(t,E_\pi;E) \,, \quad k=0,1,2,\dots,
\label{eq:7.2}
\end{equation}
as the Monte Carlo density to work with.
The suitability of this definition is further discussed over the end
of this Section.
By construction it satisfies a conservation equation like \hbox{Eq.}
(\ref{eq:6.12}), namely,
\begin{equation}
\partial_t \left[ 
\sum^A_{k=0} \tr\, {\hat \PP_k} (t,E) \right]  = 0 \,.
\label{eq:6.12a}
\end{equation}

It is clear that the evolution equation for
$\hat\rho^{(k)}(t,E_\pi;E)$ completely determines the evolution of
${\hat \PP_k} (t,E)$, however, whereas the equation for
$\hat\rho^{(k)}$ involves pions of type $k$ and $k-1$ (one-body
mechanism) that for ${\hat \PP_k}$ will depend on pions of all
classes $k'\le k$ ($N$-body quasielastic). In addition, initially
$\hat\rho^{(0)}$ is determined by $\hat\PP_0$ (cf. (\ref{eq:6.10})).
So there is an autonomous equation for the densities
${\hat \PP_k}$. The question arises how to obtain such an
equation. Simple integration over $E_\pi$ in \hbox{Eqs.}
(\ref{eq:6.11}) or (\ref{eq:6.13}) does not work. Essentially the
problem is how to invert \hbox{Eq.} (\ref{eq:7.2}). To do so I make an
{\it Ansatz\/} for the $E_\pi$ distribution of the {\it real\/} pions
in $\hat\rho^{(k)}(t,E_\pi;E)$:
\begin{equation}
\hat\rho^{(k)} (t,E_\pi;E) = \hat\rho^{(k)}_R (t, E_\pi;E) +
\hat\rho^{(k)}_V (t,E_\pi;E) \,,
\label{eq:7.3}
\end{equation}
where $\hat\rho_{R,V}$ are the distributions of real and virtual pions
in $\hat\rho$. The explicit form of these distributions is restricted
by imposing the following conditions:
\begin{itemize}
\item[1)]$\hat\rho^{(k)}_R(t,E_\pi;E)$ should be constructed out of
  ${\hat \PP_k}(t,E)$, in order to be able to invert \hbox{Eq.}
  (\ref{eq:7.2}).
\item[2)]Consistency with \hbox{Eq.} (\ref{eq:7.2}) requires that
\begin{subequations}
\label{eq:7.4}
\begin{eqnarray}
{\hat \PP_k}(t,E) &=& \int \frac{dE_\pi}{2\pi\hbar}
\hat\rho^{(k)}_R (t,E_\pi;E) \label{eq:7.4a} \\
0 &=& \int \frac{dE_\pi}{2\pi\hbar} \hat\rho^{(k)}_V (t,E_\pi;E)
\label{eq:7.4b}
\end{eqnarray}
\end{subequations}
because ${\hat \PP_k}$ is already the distribution of real pions
without the redundant $E_\pi$ dependence.
\item[3)]$\hat\rho^{(k)}_R,\hat\rho^{(k)}_V$ should be Hermitian.
\item[4)]In the classical limit, $\hat\rho^{(k)}_V$ should collapse to zero,
due to quantal interference:
\begin{subequations}
\label{eq:7.5}
\begin{eqnarray}
\hat\rho^{(k)}_V(t,E_\pi;E) & \underset{\hbar\to 0}{\longrightarrow} & 0
\label{eq:7.5a} \\
\hat\rho^{(k)}_R(t,E_\pi;E) & \underset{\hbar\to 0}{\longrightarrow} &
2\pi\hbar \delta (E_\pi - \hat H^{(\pi)}_0 ) {\hat \PP_k} (t,E) \,.
\label{eq:7.5b}
\end{eqnarray}
\end{subequations}
The $\delta$-function in \hbox{Eq.} (\ref{eq:7.5b}) follows from
(\ref{eq:7.1}) and the factor $\hat\PP$ is obtained by
normalization. In the classical limit, operators commute so there is
no conflict between (\ref{eq:7.5b}) and the point 3) above.
\item[5)]For the elastic channel we impose the constraint
\begin{equation}
 \hat\rho_V^{(k=0)} = 0
\,.
\label{eq:7.12}
\end{equation}
This defines the elastic channel pions as real. Virtual pions only
appear as a consequence of (hard) collisions, whereas the mean field
effects under the elastic evolution are regarded as soft.  Technically
this choice is needed to be able to close the equations below.

\end{itemize}

Let us see how such an {\it Ansatz\/} solves the problem. To alleviate
the notation, let us write \hbox{Eqs.} (\ref{eq:6.11}),
(\ref{eq:6.13}) in the form
\begin{subequations}
\label{eq:7.6n}
\begin{eqnarray}
\hat\rho^{(k)} &=& M^{(k,k-1)}\hat\rho^{(k-1)} \label{eq:7.6an} \\
\partial_t \hat\rho^{(k)} &=& N^{(k,k)}\hat\rho^{(k)}
+N^{(k,k-1)}\hat\rho^{(k-1)} \label{eq:7.6bn}
\end{eqnarray}
\end{subequations}
or even more compactly
\begin{subequations}
\label{eq:7.6}
\begin{eqnarray}
\rho &=& M\rho \label{eq:7.6a} \\
\partial_t \rho &=& N\rho \label{eq:7.6b}
\end{eqnarray}
\end{subequations}
where $\rho$ is a (column) vector
$\left(\hat\rho^{(0)},\hat\rho^{(1)},\ldots, \hat\rho^{(A)}\right)^T$,
and $M,N$ are matrices with respect to the index $k$ and
super-operators in Hilbert space (map operators onto operators). Also,
\hbox{Eq.} (\ref{eq:7.2}) can be written as
\begin{equation}
\PP = b\rho \,, \qquad b := \int \frac{dE_\pi}{2\pi\hbar} \,.
\label{eq:7.7}
\end{equation}
In addition, $\rho^{(k)}_R$ will be constructed out of $\PP_k$
by some linear procedure $a^{(k)}$ to be specified later:
\begin{subequations}
\begin{eqnarray}
\rho_R &=& a\PP = ab\rho \,, \label{eq:7.8} \\
\rho_V &=& (1-ab)\rho := P_V\rho \,. \label{eq:7.9}
\end{eqnarray}
\end{subequations}
$P^{(k)}_V$ is the projector onto $\hat\rho^{(k)}_V$.
Equations (\ref{eq:7.4}) then read,
\begin{equation}
ba=1 \,,\qquad bP_V = 0 \,.
\label{eq:7.10}
\end{equation}
Note that $b$ removes dependence on $E_\pi$ while $a$ creates
dependence on $E_\pi$, and $M$ increases the index $k$ by one unit.
Eq.~(\ref{eq:7.3}) then implies
\begin{equation}
\rho= \rho_R+\rho_V= a\PP+P_V\rho
= a\PP+P_VM\rho \,.
\label{eq:7.3.n}
\end{equation}
In the last step we have used the integral-like evolution equation for
expressing the short-living virtual state in terms of its source. (In
the elastic channel this is consistent due to (\ref{eq:7.12}).) This
process can be iterated until the source is a real pion and in this
way the virtual pions are integrated out. In practice this procedure
can be carried out as follows. Using previous equations and the fact
that $\partial_t$ and $b$ commute with each other, we can write the
following chain of relations:
\begin{eqnarray}
\partial_t \PP
&=& b\partial_t\rho= b\,N\rho  \nonumber \\
&=& bN\left(a\,\PP + P_V M\rho\right) \nonumber \\
&=& b\,N\,a\,\PP + b\,N\,P_V\,M\left(a\,\PP + P_VM\rho\right)\nonumber \\
&=& b\,N\,a\,\PP + b\,N\,P_V\,M\,a\,\PP + b\,N\,P_V\,M\,P_V\,M\left(a\,
\PP + P_VM\rho\right) \nonumber \\
&=& \cdots \,.
\label{eq:7.11}
\end{eqnarray}
At each step, $\rho_V$ is carried to smaller values of $k$ since it
appears with a new power of $M$. This downward recurrence ends due to
(\ref{eq:7.12}) (the elastic channel pions are real and no
inconsistency arises). In this way all virtual pions can be eliminated
and formally
\begin{equation}
\partial_t \PP = b N \left( 1 - P_V M\right)^{-1} a\PP \,.
\label{eq:7.13}
\end{equation}
This equation explicitly shows that $\PP$, as a vector, satisfies a
closed equation. Let us expand \hbox{Eq.} (\ref{eq:7.13}), with
obvious notation,
\begin{subequations}
\label{eq:7.14}
\begin{eqnarray}
\partial_t{\hat \PP_k} &=& 
\left\{ bN^{(k,k)} a^{(k)}\right\}
{\hat \PP_k} \nonumber \\ && 
+ \left\{ bN^{(k,k-1)} a^{(k-1)} + bN^{(k,k)} P^{(k)}_V M^{(k,k-1)} a^{(k-1)}
 \right\} \hat\PP_{k-1} \nonumber \\
&&  + \left\{ bN^{(k,k-1)} P^{(k-1)}_V M^{(k-1,k-2)} a^{(k-2)} \right.\nonumber \\
&&\qquad\qquad \left. + bN^{(k,k)} P^{(k)}_V M^{(k,k-1)} P^{(k-1)}_V M^{(k-1,k-2)}
 a^{(k-2)} \right\} \hat\PP_{k-2} \nonumber \\
&&  + \cdots \nonumber \\
&:=& R^{(k)} {\hat \PP_k} + Q^{(k,k-1)} \hat\PP_{k-1} + Q^{(k,k-2)}
\hat\PP_{k-2} + \cdots\,,
\label{eq:7.14a} \\
\partial_t \PP &=& R\PP + Q\PP \label{eq:7.14b}
\end{eqnarray}
\end{subequations}
where $R$ takes care of the propagation of the pion once it is
produced and $Q$ describes the quasielastic steps. (Note that in
\hbox{Eq.} (\ref{eq:1.4}) the elastic channel was included in $Q$
while here it is given by $R^{(k=0)}$.)
 
Our actual procedure has been to remove any virtual pion by explicitly
writing it in terms of the nearest real pion acting as a source for
it. As a consequence only real pions do explicitly appear, and the
hidden virtual intermediate states translate into effective $N$-body
quasielastic probabilities, \hbox{Fig.} \ref{fig:5}.

In a classical limit, each factor $P_V$ will increase the order in
$\hbar$, due to \hbox{Eq.} (\ref{eq:7.5a}), and only few terms will be
needed in (\ref{eq:7.14a}). For instance, at leading order in $\hbar$
the one-body mechanism, $Q^{(k,k-1)}$, is dominant and the effective
two-body quasielastic, $Q^{(k,k-2)}$, is the first quantum correction.

$R^{(k)}$ plays the role of an effective Hamiltonian for
${\hat \PP_k}$. Its actual form will depend on the concrete form
of $a^{(k)}$, which still has to be chosen, but using general
properties of $a^{(k)}$ we still can say something about $R^{(k)}$:
\begin{eqnarray}
R^{(k)}{\hat \PP_k} (\omega_\pi,E) &=& \left( bN^{(k,k)} a^{(k)}
\PP_k\right) (\omega_\pi,E) \nonumber \\
&=& \frac{1}{ i\hbar} \int \frac{dE_\pi}{2\pi\hbar} 
\textstyle
\left(
\hat H^{(k)}_\pi (E_\pi +\frac{1}{2}\omega_\pi)\, 
\hat\rho^{(k)}_R ( \omega_\pi, E_\pi;E) - \hat\rho^{(k)}_R (\omega_\pi,E_\pi;E)
\, \hat H^{(k)\dagger}_\pi ( E_\pi - \frac{1}{2}\omega_\pi)
\right) \nonumber \\
&=& \frac{1}{ i\hbar} \left[ \hat H^{(\pi)}_0,{\hat \PP_k} (\omega_\pi,E) \right]
\nonumber \\
&&+ \frac{1}{i\hbar} \int \frac{dE_\pi}{2\pi\hbar} 
\textstyle
\left( 
\hat V_k ( E_\pi + \frac{1}{2}\omega_\pi ) \,
\hat\rho^{(k)}_R ( \omega_\pi, E_\pi; E )
- \hat\rho^{(k)}_R ( \omega_\pi, E_\pi;E ) \,
\hat V^\dagger_k (E_\pi - \frac{1}{2}\omega_\pi)
\right)
\label{eq:7.15}
\end{eqnarray}
where $\hat V_k = \hat H^{(k)}_\pi - \hat H^{(\pi)}_0$ is the optical
potential. Further, for $\hbar\to 0$, using \hbox{Eq.} (\ref{eq:7.5b})
\begin{equation}
R^{(k)}{\hat \PP_k} (\omega_\pi,E) \approx 
\frac{1}{i\hbar} 
\textstyle
\left(
\hat H^{(k)}_\pi ( \bar{E}_\pi + \frac{1}{2}\omega_\pi) \,
{\hat \PP_k} (\omega_\pi,E) - \hat\PP_k (\omega_\pi,E) \,
\hat H^{(k)\dagger}_\pi ( \bar{E}_\pi - \frac{1}{2}\omega_\pi) 
\right)
\label{eq:7.16}
\end{equation}
where $\bar{E}_\pi = \hat H^{(\pi)}_0$ is the on-shell energy.
 
In the special case $k=0$, \hbox{Eq.} (\ref{eq:7.12}) completely
determines $a^{(k=0)}$ with the help of \hbox{Eq.} (\ref{eq:6.10})
\begin{equation}
a^{(k=0)} = 2\pi\hbar\,\delta(E-E_\pi)
\label{eq:7.17}
\end{equation}
which obviously satisfies all other requirements 1) -- 4), and the
equation $\partial_t\hat\PP_{k=0} = R^{(k=0)} \hat\PP_{k=0}$
is nothing else than \hbox{Eq.} (\ref{eq:4.11}).
 
In the general case, the super-operator $a^{(k)}$ can be chosen fairly
arbitrarily. A natural choice is
\begin{equation}
\left( a^{(k)} \hat\PP_k\right) ( t,E_\pi;E) = 
i\hbar\hat G^{(k)}_\pi (E_\pi)\, {\hat \PP_k} (t,E) 
- {\hat \PP_k} (t,E)\, i\hbar \hat G^{(k)\dagger}_\pi (E_\pi)\,, \quad k\not=0\,.
\label{eq:7.18}
\end{equation}
Clearly, this choice satisfies the requirements 1) and 3), above. The
point 2) follows from
\begin{equation}
i\hbar \hat G_\pi (E_\pi) = \int dt\,e^{iE_\pi t/\hbar} \hat U_\pi(t)
\theta (t)
\label{eq:7.19}
\end{equation}
where $\hat U_\pi(t)$ is the evolution operator, and unitarity, {\it
i.e.\/} $\hat U_\pi(t=0) = 1$. If $\hbar\to 0$, $\hat G_\pi$ and
$\hat\PP$ commute and $\hat G_\pi$ approaches the free propagator,
\begin{equation}
i\hbar\hat G^{(k)}_\pi (E_\pi) - i\hbar \hat G^{(k)\dagger}_\pi (E_\pi)
\longrightarrow 2\pi\hbar\,\delta ( E_\pi - \hat H^{(\pi)}_0)
\label{eq:7.20}
\end{equation}
which is the point 4), above. Other choices are possible, for instance
taking only $\text{Im}\, \hat G_\pi$, or taking $\hat
G^{\text{free}}_\pi$. Another possibility would be to use $E_\pi
\pm\frac{1}{2}\omega_\pi$ instead of $E_\pi$ in $\hat G_\pi$ in
(\ref{eq:7.18}) (in
frequency representation), however this choice turns out to be
inappropriate for computing $N$-body absorption processes (cf. next
Section). As compared to that, our choice of $a^{(k)}$ is the {\em
instantaneous} version since it relates $\rho^{(k)}_R$ to
$\PP_k$ at the same time.
It is noteworthy that the definition of $a^{(k)}$ has some resemblance
with the formulas invoked in the standard cascade method
\cite{Snider:1960none,Aichelin:1991xy}.

Eqs.~(\ref{eq:7.14}) are, perhaps, the main result of the paper. They
describe the evolution of the purely pionic density matrix containing
only real pions. They make explicit the $N$-body quasielastic rates
$Q^{(k,k')}$ seen by the pion as it propagates through the nucleus.
These rates are the input to be used in the cascade method. In
Appendix \ref{app:D} these formulas are trivially extended to include
absorption. They are analyzed in next Section.

It is noteworthy that the concrete choice of $a^{(k)}$ cannot affect
the evolution of ${\hat \PP_k}$, if computed to all orders, that
is, including all $N$-body mechanisms, since its definition
(\ref{eq:7.2}) is independent of $a^{(k)}$. (The right hand side of
(\ref{eq:7.13}) is actually independent of $a$.) Obviously, different
choices introduce different organizations of the series. This
ambiguity is related to that in the separation into real and virtual
pions. 
In this regard, it would be very interesting to choose $a^{(k)}$ so
that not only $bP_V^{(k)}=0$, but also $bN^{(k,k)}P_V^{(k)}=0$. This
essentially means that the real and virtual components of the density
are separately preserved under {\em elastic} evolution, i.e. in the
absence of collisions. In this case Eqs.~(\ref{eq:7.14a}) would
simplify considerably. However, it is not clear how to impose this
property or even whether it is consistent with the other requirements
set on $a^{(k)}$.  

We also note that the formulas (\ref{eq:7.14}) involve no
approximations, except that of not including crossed graphs, that is,
neglecting Bose symmetry of the ph's, as explained at the beginning of
Section \ref{sec:VI} \footnote{Another simplification has been done in
Appendix \ref{app:D} for absorption, this time just for obtaining a
simpler presentation, namely, to disregard ph self-energy graphs.}.

In Eq.~(\ref{eq:6.9}) we introduced the density matrix
$\hat\rho^{(k)}(t,E_\pi;E)$ for class $k$ pions, depending on the time
and energy of the pion and also on the conserved total energy $E$. It
can be noted that in most formulas, including (\ref{eq:6.11}),
(\ref{eq:6.13}), (\ref{eq:7.2}) and (\ref{eq:7.18}), $E$ appears only
as a parameter in the densities. An exception is (\ref{eq:7.17}),
which refers to the elastic channel, and so a separated discussion is
needed for that case.

For pions of class $k\ge 1$, $E$ is just a parameter. We could as well
introduce a new density $\hat\rho^{(k)}(t,E_\pi)$, without $E$
dependence, by integrating $\hat\rho^{(k)}(t,E_\pi;E)$ over $E$, and
rederive all equations for it. If this is done, the density of {\em
real} pions defined in (\ref{eq:7.2}) becomes ${\hat \PP_k}(t)$,
which, recalling the ``energy average'' in (\ref{eq:4.6n}),
corresponds to the standard equal-time pionic density for pions of
class $k$. In this light the integration of virtual pions is related
to the closing of a set of equations for the equal-time
densities. After the pion has scattered once or more, the initial
energy $E$ is no longer relevant and the corresponding $N$-body
reaction probabilities do not depend on $E$, a fact explicitly
verified in next Section.

On the other hand, for pions in the elastic channel $E$ is quite
relevant and coincides with $E_\pi$ (cf. (\ref{eq:6.10})).
Integrating over $E_\pi$ to obtain ${\hat \PP_k}(t,E)$ in
(\ref{eq:7.2}) merely removes this redundant energy dependence. It
should be noted that, being conserved, $E$ is a known datum in the
collision experiment, thus it does not seems advisable to remove this
information by going further and integrate over $E$ to work with the
equal-time elastic channel density matrix. Indeed, as verified in next
Section, the $N$-body reaction probabilities $Q^{(N,0)}$ and
$A^{(N,0)}$ display an explicit dependence on the pion energy.

As noted in Section \ref{sec:IV}, when a space-time Wigner form is
used (as opposed to an equal-time formulation) it is customary to use
a transport and a energy-shell constraint equation. In this work we
have used the differential-like transport equation, (\ref{eq:6.13}),
and the integral-like equation (\ref{eq:6.11}) as an equivalent set of
equations. Actually the integral-like equation contains both the
transport and the constraint equations (except for the elastic
channel) and not surprisingly this equation is responsible for putting
the particles on-shell as $\hbar$ goes to zero in the present approach
(cf. Eq~(\ref{eq:8.11})).

\section{N-body effective quasielastic and absorption mechanisms}
\label{sec:VIII}

In this Section we shall work out the consequences of the previous
scheme in the simplest cases. To simplify we shall study only nuclear
matter ($g(\bfx) = \hbox{constant} = g$ in (\ref{eq:6.14}),
$K(\bfx,\bfy) =\text{constant} = \kappa$ in (\ref{eq:D.1}), and
$\rho_\pi (\bfx,\bfp;t,E)$ independent of $\bfx$), free propagators,
locality in time, and in general lowest orders in $\hbar$. As was
shown in Section VI all these assumptions are compatible with
$\hbar\to 0$.
 
Let us study first the one-body quasielastic $Q^{(1,0)}$:
\begin{equation}
Q^{(1,0)} = bN^{(1,0)} a^{(0)} + bN^{(1,1)} P^{(1)}_V M^{(1,0)} a^{(0)} \,.
\label{eq:8.1}
\end{equation}
In leading order of $\hbar$ only the first term contributes (besides
in nuclear matter and using free Hamiltonians $N^{(k,k)}= 0$). The
result can be obtained more directly from \hbox{Eqs.} (\ref{eq:6.18}),
(\ref{eq:6.19}):
\begin{equation}
Q^{(1,0)} (\bfx,\bfp;\bfx', \bfp';E) =
\frac{1}{ \hbar^2} g^2\,\delta (\bfx-\bfx')
\,2\pi\hbar\,\delta( E - H^{(\pi)}_0(\bfp) - H_{\text{ph}} (\bfp' - \bfp)) \,.
\label{eq:8.2}
\end{equation}

The one-body absorption rate, $A^{(1,0)}$ is (the operation $P$ plays
a similar role in absorption as $N$ does in quasielastic, see Appendix
\ref{app:D} for details)
\begin{equation}
A^{(1,0)} = Pa^{(0)}
\label{eq:8.3}
\end{equation}
that can be worked out to give
\begin{equation}
A^{(1,0)} ( \bfx, \bfp; E) = \frac{\kappa^2}{ \hbar^2}
2\pi\hbar \,\delta ( E - H_{\text{ph}} (\bfp) ) \,.
\label{eq:8.4}
\end{equation}
However, in leading order in $\hbar$, $E= H^{(\pi)}_0(\bfp)$ what is
incompatible with the $\delta$-function in \hbox{Eq.} (\ref{eq:8.4})
in physical cases: the pion and the ph cannot both be on-shell and as
a consequence there is no one-body absorption in the classical limit.
 
\begin{figure}
\includegraphics[scale= 0.5]{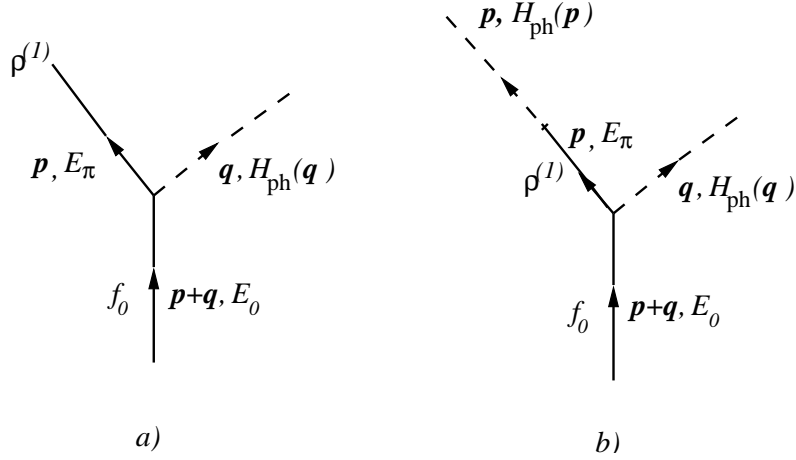}
\caption{$a$) Quasielastic step: the incoming pion and outgoing ph are
on-shell. The outgoing pion can be real or virtual. $b$) Absorption
process.}
\label{fig:6}
\end{figure}

In order to study more complicated cases, a drastic simplification in
the notation is convenient. Consider the relation $\hat\rho^{(1)} =
M^{(1,0)} a^{(0)} \hat\PP_0$ which is represented in
\hbox{Fig.}  \ref{fig:6}$a$. Using (\ref{eq:6.11}) for $M^{(1,0)}$ and
(\ref{eq:7.17}) for $a^{(0)}$, it can be written as
\begin{eqnarray}
\hat\rho^{(1)} (\omega_\pi,E_\pi;E) &=& 
\textstyle
\hat G^{(1)}_\pi (E_\pi + \frac{1}{2}\omega_\pi ) 
\tr_{\text{ph}} 
\left\{ 
2\pi\hbar\, \delta ( E - E_\pi - \hat H_{\text{ph}} ) 
\hat F^\dagger \hat\PP_0 (\omega_\pi,E) \hat F
\right\} 
\textstyle
  \hat G^{(1)\dagger}_\pi (E_\pi - \frac{1}{2}\omega_\pi)
\label{eq:6.11n}\,.
\end{eqnarray}
In what follows, we will use $E_0$ to denote the total
energy $E$ and $\PP_0(t)$ to denote (the full Wigner's form
of) $\hat\PP_0(t,E_0)$. In addition we introduce the
notations (see \hbox{Fig.} \ref{fig:6}$a$)
\begin{eqnarray}
E &:=& E_\pi - H^{(\pi)}_0(\bfp) \nonumber \\
\bar{E} &:=& E_0 - H_{\text{ph}} (\bfq) - H^{(\pi)}_0 (\bfp)
\nonumber \\
\tilde G(E) &:=& \left( E+ i\eta\right)^{-1} \,.
\label{eq:8.6}
\end{eqnarray}
Furthermore, we will focus on the $\hbar$ and energy-time dependence.
Integration over momenta, vertex operators ($\hat F$, $\hat K$),
traces, etc, will be implicit. Schematically, the relationship
$\hat\rho^{(1)} = M^{(1,0)} a^{(0)} \hat\PP_0$ will look (in
time representation, where the product of frequency functions becomes
a convolution over times)
\begin{equation}
\rho^{(1)}(t,E) = \int d\tau\,\hbar^2 \big(\tilde G\tilde G^\dagger\big)
(\tau,E)\, \PP_0 (t-\tau)\,2\pi\hbar\,\delta(E-\bar{E}) \,.
\label{eq:8.5}
\end{equation}
The total energy $E_0$ (called $E$ in previous Sections) is not
displayed. The operator $\big(\tilde G\tilde G^\dagger\big)(\tau,E)$
comes from $\hat G_\pi ( E_\pi + \frac{1}{2}\omega_\pi) \hat
G^\dagger_\pi (E_\pi - \frac{1}{2}\omega_\pi )$, namely,
\begin{equation}
\hbar^2 \big(\tilde G\tilde G^\dagger\big) (\tau,E) = \int
\frac{d\omega}{2\pi\hbar} \textstyle e^{-i\omega\tau/\hbar} \hbar^2
\tilde G( E + \frac{1}{2}\omega )\, \tilde G^\dagger ( E -
\frac{1}{2}\omega ) \,.
\label{eq:8.7}
\end{equation}
This is the kernel of the super-operator $M$ and controls how (real
and virtual) pions of class $k$ are produced out of pions of class
$k-1$ and their subsequent propagation. In terms of a standard
diagrammatic calculation (for instance in nuclear matter) this would
correspond to compute the graphs associated to the transition
amplitude between the initial and final states and then to square it
to get the transition probability, or equivalently, to compute the
self-energy graphs of the initial state and apply Cutkosky rules to
associate each cut of the graph to a contribution to the transition
probability to a concrete final state.

The exact integration over $\omega$ in (\ref{eq:8.7}) is easily
performed, but an expansion in powers of $\hbar$ is more convenient
for our purposes. To get such an expansion we start with the identity
\begin{equation}
\hbar^2 \partial_\tau \big(\tilde G \tilde G^\dagger\big) (\tau,E) =
i\hbar \big( \tilde G - \tilde G^\dagger \big) (\tau,E) \,.
\label{eq:8.8}
\end{equation}
The right-hand side is the kernel of $N$ and is a softer distribution
than $\big(\tilde G\tilde G^\dagger\big)$. Some algebra yields
\begin{eqnarray}
i\hbar\big(\tilde G - \tilde G^\dagger\big) (\tau,E) &=&
\textstyle
 -2\hbar\,\text{Im} \left[ \delta ( \tau + \frac{1}{2}i\hbar \partial_E ) 
\,\tilde G (E) \right] \nonumber \\
&=& 2\pi\hbar\,\delta(\tau)\,\delta(E) + \hbar^2 \delta'(\tau) \bar{P} \frac{1}{
E^2} + {\cal O}(\hbar^3)
\label{eq:8.9}
\end{eqnarray}
where
\begin{equation}
\bar{P} \frac{1}{ E^k} := \text{Re} \big( \tilde G(E)\big)^k = \frac{(-1)^k}{
(k-1)!}\, \frac{d^{k-1}}{ dE^{k-1}} P \frac{1}{ E}
\label{eq:8.10}
\end{equation}
is a renormalized principal value
\footnote{As a technical remark, note that for fairly general test
function spaces, the principal value of $1/x$ is well defined as a
distribution, thus, unlike $P(1/x^k)$, the construction
$\bar{P}(1/x^k)$ is also a well defined distribution for
$k=2,3,\ldots$ (finite when applied to test functions) since the
derivative of a distribution is again a distribution, defined through
by parts integration \cite{Schwartz:1957bk,Gelfand:1968:bk}.}.
Integration over $\tau$ then yields
\begin{eqnarray}
\hbar^2 \big(\tilde G\tilde G^\dagger\big) (\tau,E) &=&
\textstyle
-2\hbar\, \text{Im} \left[ \theta ( \tau + \frac{1}{2}i\hbar \partial_E ) \,
\tilde G(E)\right] \nonumber \\
&=&
2\pi\hbar\,\theta(\tau) \,\delta(E) + 
\hbar^2 \delta(\tau) \bar{P} \frac{1}{E^2} + {\cal O}(\hbar^3) \,.
\label{eq:8.11}
\end{eqnarray}
Comparing (\ref{eq:8.9}) and (\ref{eq:8.11}) we can see that
$i\hbar\big( \tilde G - \tilde G^\dagger\big)$, and so the
super-operator $N$, is instantaneous at leading order in $\hbar$
whereas $\hbar^2\big(\tilde G\tilde G^\dagger\big)$, or $M$, is
not. The non-instantaneous piece in $M$ would have been missed if we
have taken a formal classical limit neglecting the terms
$\pm\frac{1}{2}\omega$ in $\hbar^2\big(\tilde G\tilde
G^\dagger\big)$. This is incorrect because the divergence at the
on-shell pole is not integrable. Correspondingly, as we will see
subsequently, the non-instantaneous piece contributes only to real
pions and not to virtual ones. Substituting in \hbox{Eq.}
(\ref{eq:8.5}),
\begin{equation}
\rho^{(1)} (t,E) = 2\pi\hbar\,\delta(E-\bar{E})
\left( 2\pi\hbar\,\delta(\bar{E}) \partial^{-1}_t 
+ \hbar^2 \frac{1}{ \bar{E}^2} +
{\cal O}(\hbar^3)\right)\PP_0 (t)
\label{eq:8.12}
\end{equation}
where $\partial^{-1}_t = \int^t_{-\infty} dt$, and the symbol
$\bar{P}$ is implicit. We can see now what $\PP_1$,
$\rho^{(1)}_R$ and $\rho^{(1)}_V$ look like. Recalling
\begin{subequations}
\label{eq:8.13}
\begin{eqnarray}
b &=& \int \frac{dE }{ 2\pi\hbar} \,, \label{eq:8.13a} \\
a^{(k)} &=& i\hbar\left[\tilde G - \tilde G^\dagger\right]^{\text{inst.}}
(\tau,E) = 2\pi\hbar\,\delta(E)\delta(\tau),\quad k\not=0 \label{eq:8.13b}
\end{eqnarray}
\end{subequations}
where in $a^{(k)}$ only the instantaneous part should be taken due to
the missing $\pm\omega_\pi/2$ in (\ref{eq:7.18}), we have (presently
$a^{(k)}$ acts by convolution over $\tau$)
\begin{subequations}
\label{eq:8.14}
\begin{eqnarray}
\PP_1(t) &=& \left( 2\pi \hbar\,\delta(\bar{E})
\partial^{-1}_t + \hbar^2 \frac{1}{ \bar{E}^2} + {\cal O}(\hbar^3)\right)
\PP_0 (t) \label{eq:8.14a} \\
\rho^{(1)}_R (t,E) &=& 2\pi\hbar\,\delta(E) \left( 2\pi\hbar\,\delta(\bar{E})
\partial^{-1}_t + \hbar^2 \frac{1}{ \bar{E}^2} + {\cal O}(\hbar^3)\right)\PP_0(t) \label{eq:8.14b} \\
\rho^{(1)}_V (t,E) &=& 2\pi\hbar^3 \frac{1}{ \bar{E}^2} \Big( \delta(E-\bar{E}) -
\delta(E)\Big) \PP_0 (t) + {\cal O} (\hbar^4) \,.
\label{eq:8.14c}
\end{eqnarray}
\end{subequations}
We can see that in $\rho^{(1)}_R(t)$ the pions are on-shell, $E=0$,
and also that it contains a piece, that with $\partial^{-1}_t
\PP_0$, which depends on the whole previous history of the
incoming pions, {\it i.e.\/}, $\rho^{(1)}_R(t)$ contains long lasting
components indicating that the real pion at $t$ could be produced from
a quasielastic step which occurred some time ago. On the other hand,
$\rho^{(1)}_V$ does not contain such components; the number of virtual
pions depends on the instantaneous number of incoming pions indicating
that the virtual pions are short-living states.

Now we can easily compute absorption and quasielastic rates.
$Q^{(1,0)}$ will follow from
\begin{eqnarray}
Q^{(1,0)}\PP_0(t) &=& \int \frac{dE}{2\pi\hbar}
d\tau\,i\hbar\big(\tilde G-\tilde G^\dagger\big) (\tau,E)
\, 2\pi\hbar\,\delta(E-\bar{E}) \, \PP_0 (t-\tau) \nonumber \\
&=& \left( 2\pi\hbar\,\delta(\bar{E}) 
+{\cal O}(\hbar^2) \right) \PP_0 (t)
\label{eq:8.15}
\end{eqnarray}
which coincides with (\ref{eq:8.2}). Analogously, $A^{(2,1)}$ and
$A^{(2,0)}$
\begin{subequations}
\label{eq:8.16}
\begin{eqnarray}
A^{(2,1)} \PP_1 &=& Pa^{(1)} \PP_1 = \int
\frac{dE}{2\pi\hbar} 2\pi\hbar\,\delta(E-E_A) \rho^{(1)}_R (t,E)
\label{eq:8.16a} \\
A^{(2,0)} \PP_0 &=& P\rho^{(1)}_V = \int
\frac{dE}{2\pi\hbar} 2\pi\hbar\,\delta(E-E_A) \rho^{(1)}_V(t,E)
\label{eq:8.16b}
\end{eqnarray}
\end{subequations}
where $E_A := H_{\text{ph}} (\bfp) - H^{(\pi)}_0 (\bfp)$ (see
\hbox{Fig.}  \ref{fig:6}$b$). Using the expressions in (\ref{eq:8.14})
and the fact that $E_A\not=0$, as discussed for $A^{(1,0)}$, we get
\begin{subequations}
\label{eq:8.17}
\begin{eqnarray}
A^{(2,1)} &=& 0 \label{eq:8.17a} \\
A^{(2,0)} \PP_0(t) &=& 2\pi\hbar^3 \frac{1}{ \bar{E}^2}
\delta(E_A-\bar{E}) \PP_0 (t) + {\cal O}(\hbar^5) \,.
\label{eq:8.17b}
\end{eqnarray}
\end{subequations}
$Q^{(1,0)}$ and $A^{(2,0)}$ are then the lowest order quasielastic and
absorption mechanisms. Both are positive definite ($\bar{P}\frac{1}{
\bar{E}^2}$ is positive outside the pole)
\footnote{Using $E_\pi\pm\frac{1}{2}\omega_\pi$ as arguments in the
pion propagator in Eq.~(\ref{eq:7.18}) (in frequency representation),
would introduce further terms in the definition of $a^{(k)}$, (as in
(\ref{eq:8.9}) compared to (\ref{eq:8.13b})). With this alternative
definition $\rho^{(1)}_R(t,E)$ would no longer be proportional to
$\delta(E)$ and, as a consequence, $A^{(2,1)}$ would not vanish and
$A^{(2,0)}$ would not be positive at leading order (their sum,
however, would no change).}.

\begin{figure}
\includegraphics[scale= 0.5]{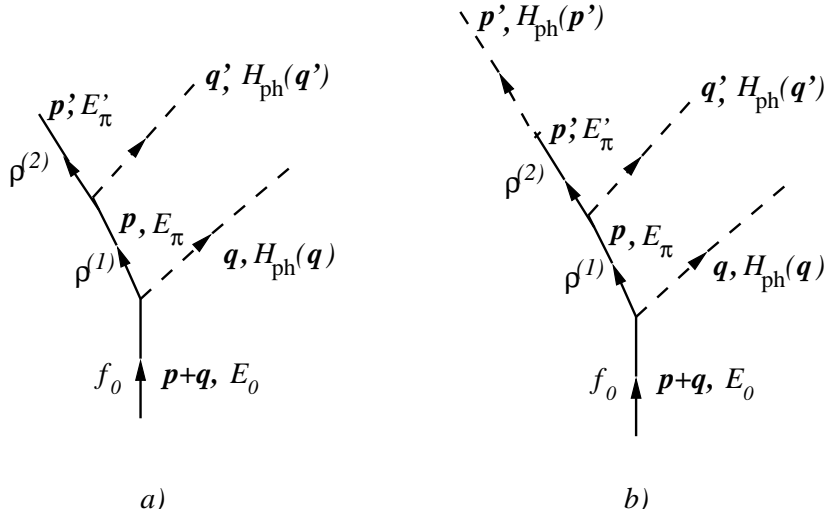}
\caption{$a$) Two-body quasielastic. $b$) Three-body absorption.}
\label{fig:7}
\end{figure}

To display further features of the present scheme we shall study the
coefficients $Q^{(2,0)}$ and $A^{(3,0)}$. Again schematically
\begin{eqnarray}
Q^{(2,1)} \PP_1(t) &=& \int \frac{dE'}{ 2\pi\hbar}d\tau'
\frac{dE}{2\pi\hbar} i\hbar\Big(\tilde G - \tilde G^\dagger\Big) (\tau',E')
\, 2\pi\hbar\,\delta(E-E'+\bar{E}') \, \rho^{(1)}_R (t-\tau',E) \nonumber \\
&=& \left( 2\pi\hbar\,\delta(\bar{E}') + {\cal O}(\hbar^2) \right) \PP_1(t)
\label{eq:8.18}
\end{eqnarray}
where $E' = E'_\pi - H^{(\pi)}_0 (\bfp')$,
$\bar{E}' = H^{(\pi)}_0(\bfp) - H^{(\pi)}_0(\bfp') - H_{\text{ph}} (\bfq')$
(see \hbox{Fig.} \ref{fig:7}$a$).

$Q^{(2,0)}$ is obtained by taking $\rho^{(1)}_V$ instead of $\rho^{(1)}_R$,
in (\ref{eq:8.18}):
\begin{equation}
Q^{(2,0)} \PP_0(t) = 2\pi\hbar^3 \frac{1}{ \bar{E}^2} 
\Big( \delta(\bar{E}' + \bar{E}) - \delta(\bar{E}') \Big) \PP_0(t) +
{\cal O}(\hbar^4) \,.
\label{eq:8.19}
\end{equation}
Note first that $Q^{(2,1)} = Q^{(1,0)}$, at least in leading order. In
general $Q^{(k+s,s)} = Q^{(k,0)}$ would be most desirable, but given
the asymmetry in the definitions of $a^{(k\not=0)}$, and $a^{(0)}$ it
might not be true for higher orders in $\hbar$. Secondly, note that
$Q^{(2,0)}$ is not positive definite. This is a consequence of
\hbox{Eq.}  (\ref{eq:7.4b}), and it was expected because it is the
first quantum correction to the quasielastic. Indeed
\begin{equation}
Q^{(2,1)}\PP_1 (t) = 2\pi\hbar\,\delta(\bar{E}')
2\pi\hbar\,\delta(\bar{E}) \partial^{-1}_t\PP_0(t) + {\cal O}(\hbar^3)
\label{eq:8.20}
\end{equation}
which shows that effectively $Q^{(2,0)}\approx \hbar Q^{(2,1)}$. It is
also interesting that as $\bar{E}\to 0$ in (\ref{eq:8.19}) and the
intermediate pion (line $\bfp$ in \hbox{Fig.} \ref{fig:7}$a$)
approaches its mass-shell, the two $\delta$-functions cancel: this is
the effect of subtracting real pions from $\rho^{(1)}$. Then, even if
the $\bar{P}\frac{1}{ \bar{E}^2}$ distribution were not already
renormalized and finite \footnote{This is the case when Cutkosky
rules \cite{Itzykson:1980bk} are directly applied in nuclear matter.}
the subtraction $\rho^{(1)} - \rho^{(1)}_R$ will produce a finite
result.
 
Similarly for absorption:
\begin{subequations}
\label{eq:8.21}
\begin{eqnarray}
A^{(3,1)} \PP_1 &=&  \int \frac{dE'}{ 2\pi\hbar}
2\pi\hbar\,\delta(E' - E'_A) \,\rho^{(2)}_{VR} 
\label{eq:8.21a} \\
A^{(3,0)}\PP_0 &=& \int \frac{dE'}{ 2\pi\hbar} 2\pi\hbar\,\delta(E'-E'_A)
\,\rho^{(2)}_{VV}
\label{eq:8.21b}
\end{eqnarray}
\end{subequations}
with $E'_A = H_{\text{ph}} (\bfp') - H^{(\pi)}_0(\bfp') \not=0$ (see
\hbox{Fig.} \ref{fig:7}$b$), and the result
\begin{subequations}
\label{eq:8.22}
\begin{eqnarray}
A^{(3,1)}\PP_1(t) &=& 2\pi \hbar^3 \frac{1}{\bar{E}'{}^2} 
\delta ( E'_A - \bar{E}' )\, \PP_1 (t) + {\cal O}(\hbar^4)
 \label{eq:8.22a} \\
A^{(3,0)} \PP_0 (t) &=& 2\pi\hbar \frac{\hbar^2}{ E^{\prime 2}_A}\
\frac{\hbar^2}{ \bar{E}^2} \Big( \delta( E'_A - \bar{E}' - \bar{E} )
- \delta(E'_A-\bar{E}') \Big) \PP_0 (t) + {\cal O}(\hbar^7) \,.
\label{eq:8.22b}
\end{eqnarray}
\end{subequations}
Once again $A^{(3,1)} = A^{(2,0)}$, also $A^{(3,0)}$ is not positive
and the $\delta$'s cancel as $\bar{E}\to 0$. Furthermore, in both
cases $Q^{(2,0)}$ and $A^{(3,0)}$ average to zero (as functions of
$\bar{E}'$), again a direct consequence of (\ref{eq:7.4b}). Although
after momentum integration $Q^{(2,0)}$ and $A^{(3,0)}$ will not be
zero, they will be small. In our formalism this is reflected in the
fact that both have higher order powers of $\hbar$. Remember that
$\hbar\to 0$ should be understood as a physical limit: it is not
$\hbar$ that is small, rather the coefficients with higher orders of
$\hbar$ are smaller if the classical limit applies.
 
Even though $Q^{(2,0)}$ and $A^{(3,0)}$ involve the propagation of
intermediate (virtual) pions, both rates are instantaneous in leading
order implying that such states last a short time. A more detailed
treatment equally shows that they run a short distance, i.e., $Q$ and
$A$ are local at leading order.

\section{Summary and conclusions}
\label{sec:IX}

In previous Sections we have reduced a complex many-body evolution
equation to something more similar to the master equation of the
cascade approach (\ref{eq:1.4}). The task has been that of rewriting
the Schr\"odinger equation and, more importantly, that of removing the
unwanted degrees of freedom, namely, the ph and the virtual pion
degrees of freedom, which are not present in the cascade-like
calculation of \hbox{Ref.} \cite{Salcedo:1988md}. We have then
obtained a set of equations for the pionic matrix density, Eqs.
(\ref{eq:7.14}), in which a classical expansion can be made naturally.

We have started by formulating the relation between the $S$-matrix and
cross section in the phase space approach, (\ref{eq:2.12}), then we
have shown that the analogous relation can be written using directly
the evolution operator, (\ref{eq:2.12n}). This relation is closer to a
cascade model since the simulation is set just to provide the
transition probability from an initial point in phase space to any
other point at large later times. Both in quantum mechanical
calculations and in cascade methods, the transition probability
(evolution operator) is obtained by solving the associated
differential equation (Schr\"odinger equation). The rest of the paper
is devoted to compute the input to be used for that equation. For this
input to be useful it has to be written in a way that connects with
cascade calculations, and we have argued that this suggests to carry
out an expansion around the classical limit. We emphasize, however,
that the validity of the final formulas (\ref{eq:6.13n1}) and
(\ref{eq:7.14}) do not rely on classical-like approximations (they are
not leading order terms of a classical expansion). They are
translations of fully quantum mechanical relations, reformulated in a
phase space form with the help of the Wigner
transformation. (Nevertheless, we recall that these formulas do not
implement indistinguishability of the ph and this can be regarded as a
classical-like feature.)

In Sections \ref{sec:III} and \ref{sec:IV} we have introduced the
necessary formalism related to the Wigner formulation. The phase space
(space-momentum) part of this formulation has been exploited already
by various authors in the literature even in the specific subject of
quantum scattering
\cite{Remler:1975fm,Remler:1979sj,Remler:1981du,Gyulassy:1983pe}. The
relevance of the {\em time-energy} form of the Wigner transformation
for many-body problems is noted in Section \ref{sec:IV} for the
elastic channel and in Section \ref{sec:V} for inelastic channels
(within a simplified model). This form is used ubiquitously later in
the paper. In Sections \ref{sec:VI} and \ref{sec:VII} we have
introduced the necessary definitions of the matrix densities until we
have pinned down the quantity, $\hat\PP$ in (\ref{eq:7.2}), that
can naturally be identified with the density of particles described by
a cascade-like method and we have also derived the evolution equations
satisfied by this quantity, (\ref{eq:7.13}). (We note that the validity
of these evolution equations holds regardless of the interpretation
given to $\hat\PP$.)  At the same time, we have systematically
studied the classical limit of the main formulas to verify their
consistency and intuitive meaning, besides, is also within a classical
expansion where they can be simple enough to be of any utility. The
classical-like expansion of (\ref{eq:7.14}) has been pursued further
in Section \ref{sec:VIII} to isolate the leading contributions to the
$N$-body quasielastic and absorption mechanisms.

Among the conclusions of this study we note:

1. The separation of the pion ``width'' in the nucleus into a
quasielastic one plus an absorptive one, is achieved by means of the
Cutkosky rules \cite{Salcedo:1988md,Salcedo:1988xd} in the
diagrammatic approach. Here, we have already been working with the
imaginary parts of the propagators from the very beginning and with
its different analytical cuts, so that no further separation is
needed.

2. We have been able to give a meaning to the concept of effective
$N$-body quasielastic and absorption probabilities. (However, beyond
lowest orders in $\hbar$ the prescription is not unique: different
choices of $a^{(k)}$ in \hbox{Eq.} (\ref{eq:7.18}) could have been
taken.) In particular, an explicit answer is given to the problem of
distinguishing three-body absorption from a quasielastic step followed
by two-body absorption. The ``genuine'' three-body absorption is given
by $A^{(3,0)}$ in (\ref{eq:8.22b}), whereas quasielastic followed by
two-body absorption is described by $Q^{(1,0)}$ in (\ref{eq:8.15}) and
$A^{(3,1)}$ in (\ref{eq:8.22a}). The rule to obtain $A^{(3,0)}$ is
essentially to subtract $A^{(3,1)}$ from the full calculation of
$\pi\to (\text{ph})^3$ obtained through a proper Feynman diagrammatic
calculation. In practice this is done by computing the relevant pion
self-energy graphs and then applying Cutkosky rules to pick up the
imaginary part, corresponding to putting the final particles on their
mass shell. The procedure of separation is further discussed in
greater detail in \cite{Salcedo:1988xd}.

3. Higher order effective quasielastic and absorption probabilities
are quantum corrections to lowest orders and they are not positive
definite. This is a direct consequence of unitarity (conservation of
number of thrown pions minus absorbed pions). A weaker condition,
namely, unitarity for large times only, would be enough but it would
require quantum interference which is prohibitive in a simulation. It
is important to note that in physical cases the {\it genuine} three
body absorption is by far dominated by the collision of the pion with
three nucleons which exchange heavier mesons, rather than by
exchanging far off-shell pions. As this heavy mesons are necessarily
virtual, the subtractions discussed here have no effect and their
contribution is positive definite, very much the same as in the two
body absorption $A^{(2,0)}$ of our model \cite{Salcedo:1988xd}. As a
matter of principle, the problem of negative probabilities can be
handled by the known method of assigning weights to the particles as
they cascade, in this case a negative weight. Unfortunately this
method introduces large statistical fluctuations. (This is the
ubiquitous negative sign problem in quantum simulations
\cite{Batrouni:1993fj,Alhassid:1994yd,Mak:1998zw}) As noted, the
concrete choice of $a^{(k)}$ in (\ref{eq:7.13}) cannot change the
result if computed to all orders (although it may affect the rate of
convergence of the expansion). Perhaps this ambiguity can be used for
reducing the importance of the negative regions in the higher order
quasielastic and absorptions rates.

4. Nuclear matter and the local density prescription appear
naturally in this scheme. Because the effective reaction rates comes
about by integrating virtual degrees of freedom, which are
quasi-local, the nuclear matter calculation results as the leading
order in a semi-classical expansion. However, one of the results of
the study in \hbox{Ref.} \cite{Salcedo:1988md} is that it is important
to take into account the finite range of the pion-nucleon interaction
which is $p$-wave. This appears in our scheme only at higher than
leading terms in $\hbar$.

5. Indistinguishability of particles has not been implemented in this
scheme. This may or may not be more than a technical problem.  It is
certainly a quantum effect. (See \cite{Padula:1990ws} for the use of
the Wigner transformation for an optimal treatment of the Bose
symmetry of pions.) We have chosen to remove the ph's and follow only
the pions as they cascade through the nucleus. It should be possible
to do a treatment without this removal, that is, with explicit ph's
(of course, the {\em virtual} ph's still have to be integrated
out). Perhaps this could have some incidence on the problem related to
the missing crossed graphs. In the context transport equations, the
problem of antisymmetry of nucleons has been successfully addressed in
extensions of quantum molecular dynamics
\cite{Feldmeier:1990st,Ono:1992uz}.

6. Although with pion-nucleus scattering in mind, the scheme is more
general and could be of interest for other problems currently dealt
with by Monte Carlo simulation methods. Furthermore, the scheme is
exemplified with a simple model, but the final equations involve
Green's functions which exist in any quantum many-body theory.

\begin{acknowledgments}
I would like to thank C. Garc\'\i a-Recio for a critical reading of
the manuscript and the Center for Theoretical Physics of the
Massachusetts Institute of Technology where the bulk of this work was
carried out some years ago. This work is supported in part by funds
provided by the Spanish DGICYT grant no. PB98-1367 and Junta de
Andaluc\'{\i}a grant no. FQM-225.
\end{acknowledgments}

\appendix

\section{}
\label{app:A}

Let us prove \hbox{Eq.} (\ref{eq:2.13}) for any free Hamiltonian of the form $\hat H_0 =
\hat H_0 (\bfp)$. From the definition of $\hat S$ we have 
$\left[ \hat H_0, \hat S\right]=0$, then for any $\hat\rho$:
\begin{equation}
e^{-it\hat H_0/\hbar} \hat S\hat\rho\hat S^\dagger\,e^{it\hat H_0/\hbar} = \hat
S\,e^{-it\hat H_0/\hbar}\hat\rho\,e^{it\hat H_0/\hbar}\hat S^\dagger  \,.
\label{eq:A.1}
\end{equation}
In Wigner's form, it implies
\begin{eqnarray}
&&\int \frac{d^3q\,d^3y}{ (2\pi\hbar)^3} e^{i\bfq (\bfx' -
\bfy)/\hbar - it \Delta H (\bfq,\bfp')/\hbar}
S( \bfy,\bfp';\bfx, \bfp) = \nonumber \\
&& \qquad \int\frac{d^3q\,d^3y}{ (2\pi\hbar)^3} 
S( \bfx',\bfp';\bfy, \bfp) e^{i\bfq (\bfy - \bfx)/\hbar - it\Delta H
(\bfq,\bfp)/\hbar}
\label{eq:A.2}
\end{eqnarray}
with $\Delta H(\bfq,\bfp) = H_0( \bfp + \frac{1}{2}\bfq ) - 
H_0(\bfp - \frac{1}{2}\bfq )$. Upon integration over $\bfx'$ and
$\bfb$ we have
\begin{eqnarray}
\int d^2b\,d^3x'S ( \bfx', \bfp'; \bfb + \bfx_\parallel,\bfp ) &=&
\int d^3x'\,d^2b \frac{dq_\parallel dy_\parallel}{ 2\pi\hbar} 
S ( \bfx', \bfp'; \bfb + \bfy_\parallel, \bfp ) 
e^{i\bfq_\parallel (\bfy_\parallel - \bfx_\parallel)/\hbar
- it\Delta H(\bfq_\parallel, \bfp)/\hbar}
\label{eq:A.3}
\end{eqnarray}
for large $t$ only small values of $\Delta H$ can survive, and $\Delta
H$ can be approximated by $\bfv (\bfp) \cdot \bfq_\parallel$
with $\bfv$ the classical velocity. Then the $\bfq_\parallel$
integration gives $\bfy_\parallel = \bfx_\parallel + \bfv t$.
But the left-hand side does not depend on $t$, then it cannot depend
on $\bfx_\parallel$ either (unless $\bfv=0$, for which case there
is no scattering at all). More technically, applying $\int
dt\,e^{i\omega t} dx_\parallel\,e^{-i\bfq' _\parallel \cdot
\bfx_\parallel}$ on both sides of (\ref{eq:A.3}) we have
\begin{equation}
0 = \left\{\delta(\omega) - \delta\big(\omega - \Delta H (\bfq_
\parallel , \bfp)\big) \right\}
\int dx_\parallel\,e^{-i\bfq_\parallel\cdot\bfx_\parallel} \int
d^2b\,d^3x' S(\bfx', \bfp'; \bfb + \bfx_\parallel,\bfp)
\label{eq:A.4}
\end{equation}
which implies (\ref{eq:2.13}) for $\bfp \not=0$.

\section{}
\label{app:B}

We want to prove \hbox{Eqs.} (\ref{eq:6.5}) and (\ref{eq:6.6}). By
iterating \hbox{Eq.} (\ref{eq:6.6}) towards lower $k$, the state
$|E_\pi, E_1, \cdots,E_k\rangle_k$ must be given by
\begin{eqnarray}
| E_\pi, E_1,\ldots,E_k\rangle_k &=& \left[ i\hbar \hat
G^{(k)}_\pi (E_\pi) i\hbar \hat G^{(k)}_{\text{ph}} (E_k) \frac{1}{i\hbar}
\hat F^\dagger \right] \nonumber \\
&&\times \left[ i\hbar \hat G^{(k-1)}_\pi (E_\pi + E_k) i\hbar\hat
G^{(k-1)}_{\text{ph}} (E_{k-1}) \frac{1}{i\hbar} \hat F^\dagger\right] \cdots \nonumber \\
&&\cdots \left[ i\hbar \hat G^{(1)}_\pi (E_\pi + E_k + \cdots + E_2 )
i\hbar \hat G^{(1)}_{\text{ph}} (E_1) \frac{1}{i\hbar}\hat F^\dagger\right] |E_\pi + E_1 +
\cdots + E_k\rangle_0 
\label{eq:B.1}
\end{eqnarray}
and for $|E\rangle_0$ we choose the state without ph particles,
$|0,E\rangle$. To illustrate the method it will sufficient to prove
\hbox{Eq.}  (\ref{eq:6.5}) just for $k=2$:
\begin{eqnarray}
|E_\pi, E_1, E_2\rangle_2 &=& \left[ i\hbar \hat G^{(2)}_\pi
(E_\pi) i\hbar\hat G^{(2)}_{\text{ph}} (E_2) \frac{1}{i\hbar}\hat F^\dagger\right] \nonumber \\
&&\times \left[ i\hbar \hat G^{(1)}_\pi (E_\pi + E_2) 
i\hbar \hat G^{(1)}_{\text{ph}} (E_1) \frac{1}{i\hbar} \hat F^\dagger\right] 
|E_\pi + E_1 + E_2\rangle_0
\label{eq:B.2} \\
|2,E\rangle' &=& \int \frac{dE_\pi}{2\pi\hbar}\, \frac{dE_1}{2\pi\hbar}\
\frac{dE_2}{ 2\pi\hbar} 2\pi\hbar\,\delta(E - E_\pi - E_1 - E_2) |E_\pi, E_1,
E_2\rangle_2 \nonumber \\
&=& \left\{ \left[ i\hbar \hat G^{(2)}_\pi \circ i\hbar\hat G^{(2)}_{\text{ph}}
\frac{1}{i\hbar}\hat F^\dagger\right] i\hbar \hat G^{(1)}_\pi\right\} \circ i\hbar \hat
G^{(1)}_{\text{ph}} \frac{1}{i\hbar} \hat F^\dagger |E\rangle_0 \,.
\label{eq:B.3}
\end{eqnarray}
Using the analytic properties of the propagator, we find
\begin{equation}
\left[ i\hbar \hat G^{(k)}_\pi \circ i\hbar \hat G^{(\ell)}_{\text{ph}}\right](E) 
= i\hbar \hat G^{(k)}_\pi \big( E - \hat H^{(\ell)}_{\text{ph}}\big) \,, 
\quad k\ge \ell \,,
\label{eq:B.4}
\end{equation}
where $\hat H^{(\ell)}_{\text{ph}}$ is the free ph Hamiltonian in the $\ell$-th
ph subspace. And similarly
\begin{eqnarray}
&& \biggl( \biggl\{ \bigl[ i\hbar\hat G^{(2)}_\pi \circ i\hbar\hat G^{(2)}_{\text{ph}} \bigr]
 i\hbar \hat G^{(1)}_\pi\biggr\} \circ i\hbar\hat G^{(1)}_{\text{ph}}
\biggr)(E) \nonumber \\
&& \qquad = i\hbar \hat G^{(2)}_\pi \big( E - \hat H^{(1)}_{\text{ph}} - \hat
H^{(2)}_{\text{ph}}\big) 
i \hbar \hat G^{(1)}_\pi \big( E - \hat H^{(1)}_{\text{ph}}\big)
\nonumber \\
&& \qquad = \left[ i \hbar \hat G^{(2)}_\pi \circ 
i\hbar\hat G^{(1)}_{\text{ph}} \circ
i\hbar \hat G^{(2)}_{\text{ph}} \right] \left[ i\hbar \hat G^{(1)}_\pi \circ
i\hbar\hat G^{(1)}_{\text{ph}} \right](E)
\label{eq:B.5}
\end{eqnarray}
where we have used the fact that $\hat G^{(k)}_\pi$ do not contain operators
related to the $k'$-th ph if $k'\le k$. This is due to our assumption that
only direct graphs are included in the propagators. The same steps in (\ref{eq:B.5})
go through when the $\hat F^\dagger$ operators are in place, as in \hbox{Eq.} (\ref{eq:B.3}),
because the last $\hat F^\dagger$ operator and $\hat G^{(1)}_{\text{ph}}$ commute: our
assumption is that each time $\hat F^\dagger$ creates the latter ph, in this case that
labeled with (2). Then
\begin{eqnarray}
|2,E\rangle^\prime &=&\left[ i\hbar\hat G^{(2)}_\pi \circ i\hbar \hat
G^{(1)}_{\text{ph}} \circ i\hbar \hat G^{(2)}_{\text{ph}} \right] \frac{1}{i\hbar}
\hat F^\dagger 
\left[ i\hbar \hat G^{(1)}_\pi \circ i\hbar\hat G^{(1)}_{\text{ph}} \right]
\frac{1}{i\hbar} \hat F^\dagger |0,E\rangle
\label{eq:B.6}
\end{eqnarray}
using \hbox{Eq.} (\ref{eq:6.1}),
\begin{equation}
|2,E\rangle' = \hat G_2 (E) \hat F^\dagger\hat  G_1 (E) \hat F^\dagger |0,E\rangle
\label{eq:B.7}
\end{equation}
and finally using \hbox{Eq.} (\ref{eq:5.9})
\begin{equation}
|2,E\rangle' = \hat G_2(E) \hat F^\dagger |1,E\rangle = |2,E\rangle \,.
\label{eq:B.8}
\end{equation}
in agreement with \hbox{Eq.} (\ref{eq:6.5}).

\section{}
\label{app:C}

In order to prove \hbox{Eq.} (\ref{eq:6.11}) let us apply the defining
operator in \hbox{Eq.} (\ref{eq:6.9}) on the recurrence
(\ref{eq:6.8}):
\begin{eqnarray}
\hat\rho^{(k)}(\omega_\pi,E_\pi;E) &=& 
{\textstyle
\hbar^2 \hat
G^{(k)}_\pi ( E_\pi + \frac{1}{2}\omega_\pi) 
}
\int \frac{dE_k}{ 2\pi\hbar}
\tr_{\text{ph}} \lim\limits_{t_k\to +\infty} \nonumber \\
&&
\times
\int \frac{d\omega_k}{2\pi\hbar} e^{-i\omega_k t_k/\hbar} 
\textstyle
\hat G_{\text{ph}} ( E_k + \frac{1}{2}\omega_k )
 \hat F^\dagger\,\hat\rho^{(k-1)} (\omega_\pi+\omega_k,E_\pi+E_k;E)
\nonumber \\
&&
\textstyle
\times \hat F 
\hat G^\dagger_{\text{ph}} ( E_k - \frac{1}{2}\omega_k )
\hat G^{(k)\dagger}_\pi ( E_\pi - \frac{1}{2}\omega_\pi ) \,.
\label{eq:C.1}
\end{eqnarray}
It is convenient to do a Fourier transformation of $\omega_\pi$ in order to
make explicit the dependence on $\omega_k$:
\begin{eqnarray}
\hat\rho^{(k)}(t_\pi,E_\pi;E) &=& 
\int \frac{d\omega_\pi}{ 2\pi\hbar} 
e^{-i\omega_\pi t_\pi/\hbar} 
{\textstyle
\hbar^2 \hat G^{(k)}_\pi ( E_\pi + \frac{1}{2}\omega_\pi) 
}
\nonumber \\ && \times
\int \frac{dE_k}{ 2\pi\hbar}
\tr_{\text{ph}} \lim\limits_{t_k\to +\infty} \int \frac{d\omega_k}{ 2\pi\hbar}
e^{-i\omega_k t_k/\hbar} dt'_\pi e^{i(\omega_\pi + \omega_k)t'_\pi/\hbar}
\nonumber \\
&&
\textstyle
\times
\hat G_{\text{ph}} ( E_k+ \frac{1}{2}\omega_k ) 
\,\hat F^\dagger\, \hat\rho^{(k-1)} ( t'_\pi, E_\pi + E_k; E ) 
\, \hat F \, \hat G^\dagger_{\text{ph}} ( E_k - \frac{1}{2}\omega_k )
\, \hat G^{(k)\dagger}_\pi ( E_\pi- \frac{1}{2}\omega_\pi ) \,.
\label{eq:C.2}
\end{eqnarray}
The structure of the $\omega_k$ integral is as follows
\begin{equation}
W= \int \frac{d\omega}{ 2\pi\hbar} e^{-i\omega t/\hbar} 
\textstyle
\hbar^2 \hat G_{\text{ph}} ( E+\frac{1}{2}\omega )
\, \hat A \, \hat G^\dagger_{\text{ph}} ( E- \frac{1}{2}\omega )
\label{eq:C.3}
\end{equation}
where $\omega=\omega_k$, $t = t_k - t'_\pi$, $E=E_k$ and $\hat A$ do not
commute with $\hat G_{\text{ph}}$.
By using the time representation of the propagators
\begin{equation}
i\hbar\hat G_{\text{ph}} (E) = \int dt\,\theta(t)\,e^{it(E-\hat H_{\text{ph}} +
i\eta)/\hbar}
\label{eq:C.4}
\end{equation}
and integrating $\omega$ \hbox{Eq.} (\ref{eq:C.3}) becomes
\begin{eqnarray}
W &=& \int dt_1 dt_2 \theta(t_1) \theta(t_2) 
\delta \left( t - \frac{t_1+t_2}{ 2}\right)
e^{it_1(E-\hat H_{\text{ph}})/\hbar} 
\hat A\,e^{-it_2(E - \hat H_{\text{ph}})/\hbar} 
\nonumber \\
&=& \int d\tau\,\theta\left( t - \left|\frac{\tau}{ 2}\right|\right)
e^{i\left( \frac{\tau}{ 2} + t\right) (E - \hat H_{\text{ph}})/\hbar} \hat
A\,e^{i\left(\frac{\tau}{ 2} - t\right) (E - \hat H_{\text{ph}})/\hbar}
\label{eq:C.5}
\end{eqnarray}
and using the cyclic property of the trace
\begin{equation}
\lim\limits_{t\to+\infty} \tr_{\text{ph}} W = \tr_{\text{ph}}
2\pi\hbar\,\delta(E-\hat H_{\text{ph}}) \hat A
\label{eq:C.6}
\end{equation}
from which \hbox{Eq.} (\ref{eq:6.11}) follows.
 
Let us now prove \hbox{Eq.} (\ref{eq:6.12}) or equivalently,
\begin{equation}
\sum_k \tr \int \frac{dE_\pi}{2\pi\hbar} \omega_\pi \hat\rho^{(k)}
(\omega_\pi, E_\pi; E) = 0 \,.
\label{eq:C.7}
\end{equation}
Our starting point is the Schr\"odinger equation (\ref{eq:6.13}).
There the part with $\hat H^{(k)}_\pi$, $\hat H^{(k)\dagger}_\pi$
contains the annihilation of the pion from $\hat\rho^{(k)}$, which go
to $\hat\rho^{(k+1)}$, while the other part contains the transition
$k-1\to k$. Then it is enough to prove that the pions that disappear
in $\hat\rho^{(k)}$ appear in $\hat\rho^{(k+1)}$: \hbox{Eq.}
(\ref{eq:C.7}) is a consequence of
\begin{eqnarray}
&&\int \frac{dE_\pi}{2\pi\hbar} \tr \left[ 
\textstyle
\hat H^{(k)}_\pi (E_\pi + \frac{1}{2}\omega_\pi )
 \hat\rho^{(k)} (\omega_\pi, E_\pi; E) -
\hat\rho^{(k)} (\omega_\pi, E_\pi; E) \hat H^{(k)\dagger}_\pi ( E_\pi -
\frac{1}{2}\omega_\pi ) \right] \nonumber \\
&& \quad
- \int \frac{dE_\pi}{2\pi\hbar}\, \frac{dE'_\pi}{ 2\pi\hbar} \tr\Biggl\{
\textstyle
2\pi\hbar\,\delta(E'_\pi - E_\pi - \hat H_{\text{ph}}) \biggl[ 
\hat G^{(k+1)}_\pi ( E_\pi+\frac{1}{2}\omega_\pi ) 
\hat F^\dagger\,\hat\rho^{(k)} (\omega_\pi,E'_\pi;E)\hat F \nonumber \\
&&
\textstyle
 \qquad
- \hat F^\dagger\,\hat\rho^{(k)} (\omega_\pi, E'_\pi;E) \hat F \,
\hat G^{(k+1)\dagger}_\pi ( E_\pi - \frac{1}{2}\omega_\pi )\biggr] \Biggr\}
= 0 \,.
\label{eq:C.8}
\end{eqnarray}
By using the cyclic property of the trace and exchanging
$E_\pi,E'_\pi$ in the second part, \hbox{Eq.} (\ref{eq:C.8}) follows
from
\begin{equation}
{\textstyle
i\hbar \big[ \hat G^{(k+1)}_\pi\circ\hat G^{(k+1)}_{\text{ph}}\big] 
( E_\pi +\frac{1}{2}\omega_\pi ) 
}
- \int \frac{dE'_\pi}{ 2\pi\hbar} 2\pi\hbar
(E_\pi - E'_\pi - \hat H^{(k+1)}_{\text{ph}} ) \, 
\textstyle
\hat G^{(k+1)}_\pi ( E'_\pi +
\frac{1}{2}\omega_\pi ) = 0
\label{eq:C.9}
\end{equation}
which is equivalent to \hbox{Eq.} (\ref{eq:B.4}).

\section{}
\label{app:D}

Here we wish to write the corresponding equations when an absorption
mechanism is included in the model of \hbox{Eq.} (\ref{eq:5.2}). We
simply add a $\pi$-ph ``vertex'' in $\hat H_I$,
\begin{equation}
\hat H_{\pi\,\text{ph}} = \hat K + \hat K^\dagger = 
\int d^3{x}\,d^3{y}\,K(\bfx,\bfy) 
\hat\phi_\pi^\dagger(\bfx)\,\hat\phi_{\text{ph}}(\bfy)
 + \text{h.c.}
\label{eq:D.1}
\end{equation}

The vertices $\hat K$, $\hat K^\dagger$ indicate that a pion can
transform into a ph and vice versa. Combined with quasielastic, this
implies that even if we start with one pion, time evolution will
produce states with no pions, and states with many pions. Below the
threshold this many-pion (more than one pion) states can only be
virtual and the exposition is greatly simplified by not considering
them. The equations generalizing (\ref{eq:5.6}) are
\begin{eqnarray}
E|\pi(\text{ph})^k\rangle &=& \hat H_0| \pi (\text{ph})^k\rangle+
\hat F^\dagger|\pi(\text{ph})^{k-1} \rangle + \hat F |\pi (\text{ph})^{k+1}\rangle 
+ \hat K |(\text{ph})^{k+1} \rangle
\nonumber \\
E|(\text{ph})^k\rangle &=& \hat H_0 |(\text{ph})^k\rangle + \hat K^\dagger|\pi(\text{ph})^{k-1} \rangle \,.
\label{eq:D.2}
\end{eqnarray}
By following the same steps as before, and again keeping only direct
graphs, it can be seen that all the formulae related to states
$|\pi(\text{ph})^k\rangle$ remain unchanged, except (\ref{eq:6.4b}):
\begin{equation}
\hat H^{(k)}_\pi = i\hbar \hat F \left[ \hat G^{(k+1)}_\pi \circ \hat
G^{(k+1)}_{\text{ph}}\right] \hat F^\dagger + \hat K\,G^{(k+1)}_{\text{ph}} \hat K^\dagger
\label{eq:D.3}
\end{equation}
the pionic Hamiltonian now contains ph's as self-energy. In this sense, the
$|\pi(\text{ph})^k\rangle$ states are autonomous, however, they act as a
source for $|(\text{ph})^{k+1}\rangle$ states:
\begin{equation}
\hat\rho^{(k)}_A (\omega_k, E_k;E) = 
\textstyle
\hat G_{\text{ph}} ( E_k + \frac{1}{2}\omega_k )
\hat K^\dagger\,\hat\rho^{(k-1)} (\omega_k, E_k; E)
\hat K\, \hat G^\dagger_{\text{ph}} ( E_k - \frac{1}{2}\omega_k )
\label{eq:D.4}
\end{equation}
where $\hat\rho^{(k)}_A$ contains the degrees of freedom of the $k$-th
ph only (recall that the other $k-1$ ph's are traced in
$\hat\rho^{(k-1)}$ which is purely pionic, and we include direct
graphs only). As usual, we can write an evolution equation for
$\hat\rho_A$, by computing $\hat G^{-1}_{\text{ph}} \hat\rho_A -
\hat\rho_A \hat G^{\dagger-1}_{\text{ph}}$. Again taking trace over ph
and integrating out $E_k$, we obtain (using the trace cyclic property)
\begin{eqnarray}
i\hbar\partial_t N^{(k)}_A (t,E) &=& \tr\int \frac{dE_k}{ 2\pi\hbar}\
\frac{d\omega_k}{ 2\pi\hbar} dt'\,e^{-i\omega_k(t-t')/\hbar} \nonumber \\
&&  
\quad \times
\textstyle
\hat K^\dagger\,\hat\rho^{(k-1)} (t',E_k;E) \,\hat K \, 
\left( 
\hat G^\dagger_{\text{ph}} ( E_k - \frac{1}{2}\omega_k ) 
- \hat G_{\text{ph}} ( E_k + \frac{1}{2}\omega_k ) \right)
\label{eq:D.5}
\end{eqnarray}
where $N^{(k)}_A(t,E)=\int {dE_k}/{ 2\pi\hbar} \,\tr\,
\hat\rho^{(k)}_A (t, E_k;E)$ denotes the number of pions absorbed by
$k$ ph until time $t$. The total number of pions absorbed is obtained
by integrating over $t$ both sides of (\ref{eq:D.5})
\begin{equation}
i\hbar N^{(k)}_A(\infty,E) = \int dt \,\tr \int \frac{dE_k}{ 2\pi\hbar} 
\hat K^\dagger\,\hat\rho^{(k-1)} (t,E_k;E)
\hat K \,2\pi i\,\delta(E_k-\hat H_{\text{ph}})
\label{eq:D.6}
\end{equation}
from which we read off the {\em effective} absorption rate
\begin{equation}
i\hbar\partial_t N^{(k)}_A(t,E) = \tr \int \frac{dE_\pi}{2\pi\hbar} 
\hat K^\dagger\,
\hat\rho^{(k-1)} (t,E_k;E) \,
\hat K \, 2\pi i\,\delta(E_k-\hat H_{\text{ph}})\,.
\label{eq:D.7}
\end{equation}
The difference between the exact, (\ref{eq:D.5}), and the effective,
(\ref{eq:D.7}), absorption rates are due to quantum fluctuations which
do not contribute to the final cross section. The simplification
occurs because the ph's are not allowed to further interact after
their creation in a quasielastic step, and so they can directly be
taken on their mass-shell (as comes out of the formula). For pions
such a simple result does not follow because the pions may always have
further quasielastic (or absorption) steps.
 
Note the odd counting in powers of $\hbar$ in \hbox{Eq.}
(\ref{eq:D.7}). However, the limit $\hbar\to 0$ will be meaningful
assuming that $\hat K$ is of order $\hbar$. The reason for this is
that actually the ph state is not elementary, as in our model, but
rather it is formed by a nucleon and its hole, and as a consequence
the field $\hat\phi_{\text{ph}}(x)$ is composite containing one loop
which gives a $\hbar$ factor to $\hat K$.

The equation similar to (\ref{eq:7.6b}) will be
\begin{equation}
\partial_t N_A = P\rho
\label{eq:D.8}
\end{equation}
where the superoperator $P$ is given in (\ref{eq:D.7}), and the
equation similar to (\ref{eq:7.13}) for absorption is
\begin{equation}
\partial_t N_A = P\left( 1 - P_V M\right)^{-1} a\PP := A\PP \,.
\label{eq:D.9}
\end{equation}
Paralleling the case of quasielastic, this formula indicates that
after integration of virtual pions there will be effective $N$-body
absorption mechanisms $A^{(k,k')}$ for the real pions.


\end{document}